\documentclass[preprint,twocolumn,10pt,pre,floatfix,a4paper]{revtex4}

\usepackage{amsmath}
\usepackage{amssymb}
\usepackage{graphicx}
\usepackage{bm}

\newcommand{\eps}{\varepsilon}
\DeclareMathOperator{\cn}{cn}
\DeclareMathOperator{\sn}{sn}
\DeclareMathOperator{\Tr}{Tr}

\def\drawgrid{
\put(-1,-2){$0$}%
\put(0,5){\vector(1,0){130}}%
\put(120,-1){$2\pi$}%
\put(63,-1){$\pi$}%
\put(128,8){$q$}%
\put(5,0){\vector(0,1){45}}%
\put(-1,44){$A$}%
\put(5,11){\line(-1,0){1}}%
\put(5,17){\line(-1,0){2}}%
\put(-1,14){$1$}%
\put(5,23){\line(-1,0){1}}%
\put(5,29){\line(-1,0){2}}%
\put(-1,26){$2$}%
\put(5,35){\line(-1,0){1}}%
\put(5,41){\line(-1,0){2}}%
\put(-1,38){$3$}%
\put( 15,5){\line(0,-1){1}}%
\put( 25,5){\line(0,-1){1}}%
\put( 35,5){\line(0,-1){1}}%
\put( 45,5){\line(0,-1){1}}%
\put( 55,5){\line(0,-1){1}}%
\put( 65,5){\line(0,-1){1}}%
\put( 75,5){\line(0,-1){1}}%
\put( 85,5){\line(0,-1){1}}%
\put( 95,5){\line(0,-1){1}}%
\put(105,5){\line(0,-1){1}}%
\put(115,5){\line(0,-1){1}}%
\put(125,5){\line(0,-1){1}}%
\put( 15,5){\dashbox{1}(0,36){}}
\put( 25,5){\dashbox{1}(0,36){}}
\put( 35,5){\dashbox{1}(0,36){}}
\put( 45,5){\dashbox{1}(0,36){}}
\put( 55,5){\dashbox{1}(0,36){}}
\put( 65,5){\dashbox{1}(0,36){}}
\put( 75,5){\dashbox{1}(0,36){}}
\put( 85,5){\dashbox{1}(0,36){}}
\put( 95,5){\dashbox{1}(0,36){}}
\put(105,5){\dashbox{1}(0,36){}}
\put(115,5){\dashbox{1}(0,36){}}
\put(125,5){\dashbox{1}(0,36){}}
}

\def\drawgridA{
\put(-1,-2){$0$}%
\put(0,5){\vector(1,0){130}}%
\put(120,-1){$2\pi$}%
\put(63,-1){$\pi$}%
\put(128,8){$q$}%
\put(5,0){\vector(0,1){45}}%
\put(-1,44){$A$}%
\put(5,14){\line(-1,0){1}}%
\put(5,23){\line(-1,0){1}}%
\put(5,32){\line(-1,0){1}}%
\put(5,41){\line(-1,0){2}}%
\put(-1,38){$1$}%
\put( 15,5){\line(0,-1){1}}%
\put( 25,5){\line(0,-1){1}}%
\put( 35,5){\line(0,-1){1}}%
\put( 45,5){\line(0,-1){1}}%
\put( 55,5){\line(0,-1){1}}%
\put( 65,5){\line(0,-1){1}}%
\put( 75,5){\line(0,-1){1}}%
\put( 85,5){\line(0,-1){1}}%
\put( 95,5){\line(0,-1){1}}%
\put(105,5){\line(0,-1){1}}%
\put(115,5){\line(0,-1){1}}%
\put(125,5){\line(0,-1){1}}%
\put( 15,5){\dashbox{1}(0,36){}}
\put( 25,5){\dashbox{1}(0,36){}}
\put( 35,5){\dashbox{1}(0,36){}}
\put( 45,5){\dashbox{1}(0,36){}}
\put( 55,5){\dashbox{1}(0,36){}}
\put( 65,5){\dashbox{1}(0,36){}}
\put( 75,5){\dashbox{1}(0,36){}}
\put( 85,5){\dashbox{1}(0,36){}}
\put( 95,5){\dashbox{1}(0,36){}}
\put(105,5){\dashbox{1}(0,36){}}
\put(115,5){\dashbox{1}(0,36){}}
\put(125,5){\dashbox{1}(0,36){}}
}

\def\drawgridB{
\put(-1,-2){$0$}%
\put(0,5){\vector(1,0){130}}%
\put(120,-1){$2\pi$}%
\put(63,-1){$\pi$}%
\put(128,8){$q$}%
\put(5,0){\vector(0,1){45}}%
\put(-1,44){$A$}%
\put(5,14){\line(-1,0){1}}%
\put(5,23){\line(-1,0){2}}%
\put(-1,20){$1$}%
\put(5,32){\line(-1,0){1}}%
\put(5,41){\line(-1,0){2}}%
\put(-1,38){$2$}%
\put( 15,5){\line(0,-1){1}}%
\put( 25,5){\line(0,-1){1}}%
\put( 35,5){\line(0,-1){1}}%
\put( 45,5){\line(0,-1){1}}%
\put( 55,5){\line(0,-1){1}}%
\put( 65,5){\line(0,-1){1}}%
\put( 75,5){\line(0,-1){1}}%
\put( 85,5){\line(0,-1){1}}%
\put( 95,5){\line(0,-1){1}}%
\put(105,5){\line(0,-1){1}}%
\put(115,5){\line(0,-1){1}}%
\put(125,5){\line(0,-1){1}}%
\put( 15,5){\dashbox{1}(0,36){}}
\put( 25,5){\dashbox{1}(0,36){}}
\put( 35,5){\dashbox{1}(0,36){}}
\put( 45,5){\dashbox{1}(0,36){}}
\put( 55,5){\dashbox{1}(0,36){}}
\put( 65,5){\dashbox{1}(0,36){}}
\put( 75,5){\dashbox{1}(0,36){}}
\put( 85,5){\dashbox{1}(0,36){}}
\put( 95,5){\dashbox{1}(0,36){}}
\put(105,5){\dashbox{1}(0,36){}}
\put(115,5){\dashbox{1}(0,36){}}
\put(125,5){\dashbox{1}(0,36){}}
}

\begin{document}
\title{Stability of Nonlinear Normal Modes in the FPU-$\beta$ Chain in the Thermodynamic Limit}

\author{G.~M.~Chechin}
\email{gchechin@gmail.com}
\author{D.~S.~Ryabov}
\email{dryabov@yandex.ru}
\affiliation{Institute of Physics, Southern Federal University, Rostov-on-Don, Russia}

\date{\today}

\begin{abstract}
All possible symmetry-determined nonlinear normal modes (also called by simple periodic orbits, one-mode solutions etc.) in both hard and soft Fermi-Pasta-Ulam-$\beta$ chains are discussed. A general method for studying their stability in the thermodynamic limit, as well as its application for each of the above nonlinear normal modes are presented.
\end{abstract}


\maketitle


\section{Introduction}

This paper is devoted to studying the stability of dynamical objects which are called by very different terms such as one-mode solutions (OMSs)~\cite{PoggiRuffo1997,LeoLeo2007}, simple periodic orbits (SPOs)~\cite{Bountis2006}, low-dimensional solutions~\cite{Shinohara}, one-dimensional bushes~\cite{FPU1,FPU2} etc. Below we refer to them as \textit{nonlinear normal modes} (NNMs). Let us comment on this terminology.

The concept of similar nonlinear normal modes was developed by Rosenberg many years ago~\cite{Rosenberg}. Each NNM represents a periodic vibrational regime in the conservative $N$-particle mechanical system for which the displacement $x_i(t)$ of every particle is proportional to the displacement of an arbitrary chosen particle, say, the first particle [$x_1(t)$] at any instant $t$:
\begin{equation}\label{eqch1}
x_i(t)=c_i\cdot x_1(t),
\end{equation}
where $\{c_1=1,\ c_2,\ c_3,\ \ldots,\ c_N\}$ are constant coefficients.\footnote{We give the definition of NNMs for the one-dimensional case since below the longitudinal vibrations of the Fermi-Pasta-Ulam chains are only considered.}

Note that convenient \textit{linear normal modes} (LNMs) also satisfy Eq.~(\ref{eqch1}) since, for any such mode, one can write
\begin{equation}\label{eqch2}
x_i(t)=a_i\cdot\sin(\omega t+\phi_0),\quad i=1,\ldots,N,
\end{equation}
where $a_i$ are constant amplitudes of individual particles, while $\omega$ and $\phi_0$ are the frequency and initial phase of the considered mode.

As a rule, NNMs can exist in the mechanical systems with rather specific interparticle interactions, for example, in systems whose potential energy represents a \textit{homogeneous} function with respect to all its arguments. However, in some cases, the existence of NNMs is caused by certain symmetry-related reasons. We refer to such dynamical objects as \textit{symmetry-determined} NNMs.

In~\cite{ChechinStokes}, we have found all symmetry-determined NNMs in all $N$-particle mechanical systems with any of 230 space groups. This proved to be possible due to the group-theoretical methods developed in~\cite{DAN1,DAN2,PhysD98} for constructing \textit{bushes} of vibrational modes.

At this point, it is worth to comment on the concept of bushes of modes introduced in~\cite{DAN1} (the theory of these dynamical objects can be found in~\cite{PhysD98,Columbus}).

In rigorous mathematical sense, they represent symmetry-determined \textit{invariant manifolds} decomposed into the basis vectors of \textit{irreducible representations} of the symmetry group characterizing the considered mechanical system (``parent'' group).

Because of the specific subject of the present paper, it is sufficient to consider only bushes of vibrational modes in nonlinear monoatomic chains. Such bushes have been discussed in~\cite{FPU1,FPU2}. Let us reproduce here some ideas and results from these papers.

Every bush B$[G]$ describes a certain vibrational regime, and some specific \textit{pattern} of instantaneous displacements of all the particles of the mechanical system corresponds to it. In turn, this pattern is characterized by a symmetry group $G$ (in particular, such group can be trivial) which is a \textit{subgroup} of the symmetry group $G_0$ of the mechanical system in its equilibrium state.

For example, let us consider the \textit{two-dimensional} bush B$[a^4,i]$ in the monoatomic chain with periodic boundary conditions whose displacement pattern $\vec X(t)$ can be written as follows
\begin{equation}\label{eqch3}
\begin{split}
\vec X(t)=&\{x_1(t),x_2(t),-x_2(t),-x_1(t)~|~x_1(t),x_2(t),\\
&\quad-x_2(t),-x_1(t)~|~\ldots\}.
\end{split}
\end{equation}
This pattern is determined by two time-dependent functions $x_1(t)$, $x_2(t)$, and the corresponding \textit{vibrational state} of the $N$-particle chain is fully described by displacements inside the \textit{primitive cell}, which is four time larger than that of the equilibrium state.

We will refer to the ratio of the primitive cell size of the vibrational state to that of the equilibrium state as \textit{multiplication number} ($MN$) and, therefore, for the pattern~(\ref{eqch3}), one can write $MN=4$.

The symmetry group ${G=[a^4,i]}$ of the bush B$[a^4,i]$ is determined by two \textit{generators}: the translation ($a^4$) by four lattice spacing ($a$) and the inversion ($i$) with respect to the center of the chain (note that the condition ${N\bmod 4=0}$ must hold for existence of such bush).

If we decompose the displacement pattern~(\ref{eqch3}) into the linear normal coordinates\footnote{Here we use the linear normal modes in the form that was used in~\cite{PoggiRuffo1997}.}
\begin{align}
\vec\Psi_j=&\left\{\left.\frac{1}{\sqrt{N}}\left[\sin\left(\frac{2\pi j}{N}n\right)+\cos\left(\frac{2\pi j}{N}n\right)\right]\right|n=1..N\right\}\nonumber\\
&(j=0..N-1),\label{eqch10}
\end{align}
we get the following form of the bush B$[a^4,i]$ in the \textit{modal space}:
\begin{equation}\label{eqch12}
\vec X(t)=\mu(t)\vec\Psi_{N/2}+\nu(t)\vec\Psi_{3N/4},
\end{equation}
where
\begin{equation}\label{eqch13}
\vec\Psi_{N/2}=\frac{1}{\sqrt{N}}\{-1,1~|~-1,1~|~-1,1~|~-1,1~|~\ldots\},
\end{equation}
\begin{equation}\label{eqch14}
\vec\Psi_{3N/4}=\frac{1}{\sqrt{N}}\{-1,-1,1,1~|~-1,-1,1,1~|~\ldots\},
\end{equation}
while $\mu(t)$ and $\nu(t)$ are time-dependent coefficients in front of the normal coordinates $\vec\Psi_{N/2}$ and~$\vec\Psi_{3N/4}$.

Thus, only $m=2$ normal coordinates from the full set~(\ref{eqch10}) contribute to the ``configuration vector'' $\vec X(t)$ corresponding to the given bush and we will refer to $m$ as the \textit{bush dimension}.

In~\cite{FPU2}, we developed a simple crystallographic method for obtaining the displacement pattern~$\vec X(t)$ for any subgroup~$G$ of the parent group~$G_0$. Using this method one can obtain bushes of different dimensions for an arbitrary nonlinear chain. The \textit{one-dimensional bushes} (${m=1}$) represent symmetry-determined nonlinear normal modes. The displacement pattern~$\vec X(t)$ corresponding to a given NNM depends on only one (time-periodic) function $\nu(t)$:
\begin{equation}\label{eqch20}
\vec X(t)=\nu(t)\cdot\vec c,
\end{equation}
where $\vec c$ is a constant vector, which is formed by the coefficients $c_i$ (${i=1..N}$) from Eq.~(\ref{eqch1}), while the function $\nu(t)$ satisfies a certain differential equation. This so-called ``governing'' equation can be obtained by substitution of the ansatz~(\ref{eqch20}) into the dynamical equations of the considered chain.

In some sense, the concept of bushes of vibrational modes can be considered as a certain \textit{generalization} of the notion of NNMs by Rosenberg. Indeed, if we substitute the ansatz~(\ref{eqch12}) into dynamical equations of the chain, we obviously get two ``governing'' equations for the functions $\nu(t)$ and $\mu(t)$, that determines the above-discussed two-dimensional bush (note that, in contrast to a NNM, such dynamical object describes, in general, a \textit{quasiperiodic} motion). Finally, one can conclude that $m$-dimensional bush is determined by $m$ time-dependent functions for which $m$ governing differential equations can be obtained from the dynamical equations of the considered mechanical system.

Let us emphasize that bushes of modes represent a new type of \textit{exact} excitations in nonlinear systems with discrete symmetries and the excitation energy proves to be trapped in a given bush for the case of Hamiltonian systems.

It is very important to emphasize that there exist only a \textit{finite number} of vibrational bushes of any fixed dimension in every $N$-particle mechanical system. As a consequence, there is sufficiently small number of NNMs (one-dimensional bushes) in the FPU chains (three NNMs for the FPU-$\alpha$ model and six---for the FPU-$\beta$ model).

All possible one-dimensional bushes are explicitly listed in our papers~\cite{FPU1,FPU2} (see also~\cite{Rink}).

The stability of some NNMs in the FPU chains has been studied in~\cite{Budinsky1983,BermanKolovskij1984,SanduskyPage1994,Flach1996,Dauxois1997,PoggiRuffo1997,Shinohara,Yoshimura2004,Dauxois2005,Bountis2006,Bountis2006a,LeoLeo2007,LeoLeo2011} by numerical and analytical methods. Let us comment explicitly on the recent papers~\cite{Bountis2006,LeoLeo2007,LeoLeo2011}.

In~\cite{Bountis2006,Bountis2006a}, T.~Bountis and coworkers have investigated the destabilization thresholds ($E_{1u}$ and $E_{2u}$) of two nonlinear normal modes which they call SPO-1 and SPO-2 (simple periodic orbits) by numerical methods. The authors of the above papers try to reveal some relations between the destabilization thresholds $E_{1u}$, $E_{2u}$ and the origin of the weak chaos in connection with the famous Fermi-Pasta-Ulam problem of the energy equipartition between different modes. In particular, they conclude that the main role in the weak chaos appearance in the thermodynamic limit ($N\rightarrow\infty$) plays SPO-2, because $E_{2u}\sim\frac{1}{N^2}$, $E_{1u}\sim\frac{1}{N}$, and, therefore, $E_{2u}<E_{1u}$.

However, there are some other SPOs in the FPU-$\beta$ chain and one can be interested in their role in the origin of the weak chaos in the thermodynamic limit. Some comments are appropriate at this point.

According to Lyapunov~\cite{Lyapunov}, some strictly periodic orbits for nonlinear systems can be obtained from the linear normal modes (which are introduced in the harmonic approximation) by continuation with respect to a parameter characterizing the strength of nonlinearity.\footnote{This procedure proves to be possible if frequencies of the normal modes are rationally independent.} From this point of view, there exist $N$ different SPOs for longitudinal vibrations of an $N$ particle monoatomic chain. However, only few of the modes, constructed in such a way, possess an \textit{identical} time dependence of the displacements of all the particles. More exactly, only few of the Lyapunov modes can be written in the form~(\ref{eqch20}) implying a separation of time and space variables that is typical for the Rosenberg nonlinear normal modes. Indeed, in general case, $x_i(t)=\nu_i(t)\cdot c_i$ where $\nu_i(t)$ ($i=1,\ldots,N$) are \textit{different} functions of time with identical periods.

Note that in the present paper we consider only \textit{extended} SPOs, but the same problem there exists for \textit{localized} periodic modes (discrete breathers) and we have discussed it in detail in~\cite{SoundVibr}.

As far as we aware, all periodic solutions in monoatomic chains that have been studied up to now (see the above cited papers~\cite{PoggiRuffo1997,LeoLeo2007,Bountis2006,Shinohara} and references therein) belong namely to the class of the Rosenberg nonlinear normal modes determined by Eq.~(\ref{eqch20}). Moreover, the spatial profiles $\vec c=\{c_1,c_2,\ldots,c_N\}$ of these modes possess certain symmetry properties. In particular, every such mode can be characterized by a multiplication number ($MN$) determining the enlargement of the primitive cell of the vibrational state in comparison with that of the equilibrium state. As was already noted, we refer to these modes as symmetry-determined NNMs and there exist only finite number of such modes (even for the case $N\rightarrow\infty$!) for each nonlinear chain~\cite{FPU1,FPU2,Rink}.

Above considered SPO-1, SPO-2 and the well-known $\pi$-mode (zone boundary mode) represent NNMs with multiplication numbers 4, 3, and 2, respectively. However, among \textit{six} symmetry-determined NNMs in the FPU-$\beta$ chain~\cite{FPU2} there exist another three NNMs with $MN=3$, $MN=4$ and $MN=6$.

The stability properties of the second NNM with $MN=4$ were studied by M.~Leo and R.A.~Leo in~\cite{LeoLeo2007}. The stability of this mode was investigated in the thermodynamic limit by both numerical and analytical methods.

The stability diagrams for all the nonlinear normal modes in the FPU-$\beta$ chain, as well as for the FPU-$\alpha$ chain, can be found in our paper~\cite{FPU2}. With the aid of these diagrams, one may reveal many stability properties of NNMs for an arbitrary $N$, in particular, for the thermodynamic limit ($N\rightarrow\infty$). Note that these diagrams were obtained numerically.

In this paper, we present some \textit{analytical} results for the stability properties of all NNMs in the FPU-$\beta$ chain in the \textit{thermodynamic} limit ($N\rightarrow\infty$). We also compare our results with those by different authors when it is possible.

In Sec.~2, we consider all the possible symmetry-determined nonlinear normal modes in the FPU-$\beta$ chain. In Sec.~3, the stability diagrams for these NNMs are discussed. In Sec.~4, the analytical method for studying the stability of NNMs in the thermodynamic limit is presented. In Sec.~5, we list our results on the stability properties for every NNM.

\section{Nonlinear normal modes in the FPU-$\beta$ chain}

As was already mentioned, there exists only \textit{finite} number of symmetry-determined NNMs in any monoatomic chain. Every NNM corresponds to a certain \textit{subgroup} of the symmetry group of the chain dynamical equations. The difference in the number of nonlinear normal modes for the FPU-$\alpha$ chain (three NNMs) and the FPU-$\beta$ chain (six NNMs) is associated with the fact that the symmetry group of the FPU-$\beta$ chain dynamical equations is higher than that of the FPU-$\alpha$ chain~\cite{FPU1,FPU2,Rink}.

In~\cite{FPU2}, we have investigated the stability of all NNMs both in the FPU-$\alpha$ and FPU-$\beta$ chains (for the case $\beta>0$) by numerical methods. Let us comment on the main idea of this investigation.

Following the standard method of the linear stability analysis, we linearize the FPU-$\beta$ dynamical equations near a given NNM and get the linearized system in the form $\ddot{\vec\delta}=J(t)\,\vec\delta$, where $\vec\delta(t)$ represents the infinitesimal perturbation vector, while $J(t)$ is the Jacobian matrix of the original system of nonlinear differential equations. Thus, we obtain $N$ linear differential equations with time-periodic coefficients depending on the considered NNM. Then the Floquet method can be applied for studying the stability of the zero solution of the system~$\ddot{\vec\delta}=J(t)\,\vec\delta$.

However, such straightforward way for the stability analyzing becomes practically impossible for ${N\rightarrow\infty}$. In~\cite{Zhukov}, the general group-theoretical method has been developed for splitting (decomposition) of the original system $\ddot{\vec\delta}=J(t)\,\vec\delta$ of $N$ linear differential equations into certain subsystems of sufficiently small dimensions ${N_j\ll N}$. For the FPU-$\beta$ chain, these dimensions do not exceed three (see below). Then we have applied the Floquet method for such subsystems of small dimensions. Moreover, proceeding in this manner, one can reveal those subsets of the vibrational modes, which are responsible for the loss of stability of the considered NNM. As a consequence of this approach, it proves to be possible to construct very transparent diagrams, which demonstrate explicitly stability properties of each FPU nonlinear normal mode~\cite{FPU2}.

The FPU-$\beta$ model represents a chain of unit masses coupled with each other by the appropriate nonlinear springs. The dynamical equations describing longitudinal vibrations of the FPU-$\beta$ chain can be written in the form
\begin{equation}\label{eqch35}
\ddot x_i=f(x_{i+1}-x_i)-f(x_i-x_{i-1}),\quad i=1..N,
\end{equation}
where $x_i(t)$ is the displacement of the $i$th particle from its equilibrium state at the instant $t$, while the force $f(\Delta x)$ depends on the spring deformation $\Delta x$ as
\begin{equation}\label{eqch36}
f(\Delta x)=\Delta x+\beta(\Delta x)^3.
\end{equation}
The periodic boundary condition is assumed to hold:
\begin{equation}\label{eqch37}
x_{N+1}(t)\equiv x_1(t),\quad x_0(t)\equiv x_N(t).
\end{equation}

Let us mention some results of the paper~\cite{FPU2}, which are necessary for our further discussions.

Every NNM in the FPU-$\beta$ chain can be written as follows [see Eq.~(\ref{eqch20})]:
\[
\vec X(t)=\nu(t)\cdot\vec c,
\]
where $\nu(t)$ satisfies the Duffing equation
\begin{equation}\label{eqch38}
\ddot\nu+\omega^2\nu+\gamma\frac{\beta}{N}\nu^3=0
\end{equation}
with different values $\omega$ and $\gamma$ for different NNMs. The function~$\nu(t)$ describes the time-evolution of a given NNM, while the $N$-dimensional vector~$\vec c$ determines the pattern of the displacements of all particles of the chain.

Below, we list all possible NNMs in the FPU-$\beta$ chain.

\begin{enumerate}
\item B$[a^2,i]$:
\begin{equation}\label{eqch_60}
\begin{array}{c}
\vec c=\frac{1}{\sqrt{N}}\{1,-1~|~1,-1~|~1,-1~|~\ldots\}, \\
\omega^2=4,\quad\gamma=16\quad(N\bmod 2=0).
\end{array}
\end{equation}
This is a boundary zone mode or $\pi$-mode.

\item B$[a^3,i]$:
\begin{equation}\label{eqch_61}
\begin{array}{c}
\vec c=\frac{3}{\sqrt{6N}}\{1,0,-1~|~1,0,-1~|~1,0,-1~|~\ldots\}, \\
\omega^2=3,\quad\gamma=\frac{27}{2}\quad(N\bmod 3=0).
\end{array}
\end{equation}
There exist three ``dynamical domains'' of this NNM (see below).

\item B$[a^4,ai]$:
\begin{equation}\label{eqch_62}
\begin{array}{c}
\vec c=\frac{2}{\sqrt{2N}}\{0,1,0,-1~|~0,1,0,-1~|~0,1,0,-1~|~\ldots\}, \\
\omega^2=2,\quad\gamma=4\quad(N\bmod 4=0).
\end{array}
\end{equation}
There exist two dynamical domains of this NNM.

\item B$[a^3,iu]$:
\begin{equation}\label{eqch_63}
\begin{array}{c}
\vec c=\frac{1}{\sqrt{2N}}\{1,-2,1~|~1,-2,1~|~1,-2,1~|~\ldots\}, \\
\omega^2=3,\quad\gamma=\frac{27}{2}\quad(N\bmod 3=0).
\end{array}
\end{equation}
There exist three dynamical domains of this NNM.

\item B$[a^4,iu]$:
\begin{equation}\label{eqch_64}
\begin{array}{c}
\vec c=\frac{1}{\sqrt{N}}\{1,-1,-1,1~|~1,-1,-1,1~|~1,-1,-1,1~|~\ldots\}, \\
\omega^2=2,\quad\gamma=8\quad(N\bmod 4=0).
\end{array}
\end{equation}
There exist two dynamical domains of this NNM.

\item B$[a^3u,ai]$:
\begin{equation}\label{eqch_65}
\begin{array}{c}
\vec c=\frac{3}{\sqrt{6N}}\{0,1,1,0,-1,-1~|~0,1,1,0,-1,-1~|~\ldots\}, \\
\omega^2=1,\quad\gamma=\frac{3}{2}\quad(N\bmod 6=0).
\end{array}
\end{equation}
There exist three dynamical domains of this NNM.
\end{enumerate}

Let us comment on the above listed NNMs in the FPU-$\beta$ chain.

Every NNM, denoted by the symbol B$[G]$, is characterized by the corresponding symmetry group $G$, that represents a certain subgroup of the symmetry group ${G_0=[a,i,u]}$ of the FPU-$\beta$ dynamical equations~(\ref{eqch35},\ref{eqch36}). We determine every such group by the set of its generators using the following notations:

\begin{description}
\item[$a$]--- the translation of the chain by one lattice spacing,
\item[$i$~]--- the inversion with respect to the center of the chain,
\item[$u$]--- the operator, that changes signs of the displacements of all particles without any their transposition.
\end{description}

The symmetry group ${G_0=[a,i,u]}$ of the FPU-$\beta$ dynamical equations is described by \textit{three} generators ($a$, $i$, and $u$). The corresponding transformations $a$, $i$ and $u$ of $N$-dimensional vectors ${\vec x=\{x_1,x_2,\ldots,x_N\}}$ \textit{do not} change the dynamical equations~(\ref{eqch35},\ref{eqch36}) of the FPU-$\beta$ chain.

All the above listed groups of NNMs are fully described by only \textit{two} generators, but these generators can be written as some \textit{products} of the generators $a$, $i$, and $u$ of the group $G_0$. For example, $a^2$, $a^3$, $a^4$ are translations of the chain by two, three and four lattice spacings, respectively. The transformation $ai$ means that we must perform the inversion of the displacement pattern with respect to the chain center and then translate it by one lattice spacing.

Note that transformations $a$ and $i$ do not commute:
\begin{equation}\label{eqch40}
ia=a^{-1}i,\quad\text{or}\quad ia=a^{N-1}i
\end{equation}
[the relation $a^{-1}=a^{N-1}$ holds because of the periodic boundary condition~(\ref{eqch37})]. On the other hand, the transformation $u$ does commute with both $a$ and $i$ transformations:
\begin{equation}\label{eqch41}
ua=au,\quad ui=iu.
\end{equation}

The transformation $a^2u$ means that we must change signs of all displacements and then translate the displacement pattern $\vec X$ by two lattice spacings.

Some simple examples are worth mentioning at this point. For the chain with $N=6$ particles, we can write the following relations:
\[
\begin{array}{l}
a\,\{x_1,x_2,x_3,x_4,x_5,x_6\}=\{x_2,x_3,x_4,x_5,x_6,x_1\},\\
i\,\{x_1,x_2,x_3,x_4,x_5,x_6\}=\\
\quad\quad\quad\quad\quad\quad\{-x_6,-x_5,-x_4,-x_3,-x_2,-x_1\},\\
iu\,\{x_1,x_2,x_3,x_4,x_5,x_6\}=\{x_6,x_5,x_4,x_3,x_2,x_1\},\\
a^2i\,\{x_1,x_2,x_3,x_4,x_5,x_6\}=\\
\quad\quad\quad\quad\quad\quad\{-x_4,-x_3,-x_2,-x_1,-x_6,-x_5\},\\
a^2u\,\{x_1,x_2,x_3,x_4,x_5,x_6\}=\\
\quad\quad\quad\quad\quad\quad\{-x_3,-x_4,-x_5,-x_6,-x_1,-x_2\},\\
\text{etc.}
\end{array}
\]

The displacement pattern corresponding to a given NNM can be obtained as \textit{invariant vector} of its symmetry group ${G\subset G_0}$. For example, let us obtain the displacement pattern for the NNM with $G=[a^4,ai]$ [see Eq.~(\ref{eqch_62})]. For simplicity, we demonstrate the method for obtaining displacement patterns with the case $N=8$. Let
\[
\vec X=\{x_1,x_2,x_3,x_4,x_5,x_6,x_7,x_8\}
\]
where $x_i$ ($i=1..8$) are arbitrary displacements of eight particles of the chain. The vector $\vec X$ must be invariant with respect to the action of our two generators $a^4$ and $ai$ of the symmetry group of the considered NNM:
\begin{equation}\label{eqch65}
a^4\vec X=\vec X,\quad ai\vec X=\vec X.
\end{equation}
The former equation is reduced to the following form:
\begin{align*}
a^4\vec X=\{x_5,x_6,x_7,x_8~|~x_1,x_2,x_3,x_4\}=\\
\{x_1,x_2,x_3,x_4~|~x_5,x_6,x_7,x_8\},
\end{align*}
from which we conclude that
\[
x_5=x_1,\quad x_6=x_2,\quad x_7=x_3,\quad x_8=x_4.
\]
This displacement pattern is formed by two primitive cells whose size four times larger than that of the FPU-$\beta$ chain in its equilibrium state. The sets of the displacements in both cells are identical:
\begin{equation}\label{eqch66}
\vec X=\{x_1,x_2,x_3,x_4~|~x_1,x_2,x_3,x_4\}.
\end{equation}

Now let us take into account the second equation~(\ref{eqch65}). Acting on the vector~(\ref{eqch66}) by $ai$, we obtain
\begin{align*}
ai\vec
X=a\{-x_4,-x_3,-x_2,-x_1~|~-x_4,-x_3,-x_2,-x_1\}=\\
=\{-x_3,-x_2,-x_1,-x_4~|~-x_3,-x_2,-x_1,-x_4\}.
\end{align*}
Then using the equation $ai\vec X=\vec X$, we get
\[
x_1=-x_3,\quad x_2=-x_2=0,\quad x_4=-x_4=0.
\]
Thus, the invariant (under the action of the group ${G=[a^4,ai]}$) vector $\vec X$ depends on \textit{only one} arbitrary parameter, which we denote by $x$:
\begin{equation}\label{eqch60}
\vec X=\{x,0,-x,0~|~x,0,-x,0\}.
\end{equation}
(Note that this vector being invariant with respect to generators of the group ${G=[a^4,ai]}$ will automatically be invariant relative to all its other elements). Then the NNM corresponding to the invariant vector~(\ref{eqch60}) can be written as follows
\begin{align}
\vec
X_{NNM}(t)=\{\nu(t),0,-\nu(t),0~|~\nu(t),0,-\nu(t),0\}=\nonumber\\
\nu(t)\cdot\{1,0,-1,0~|~1,0,-1,0\}.\label{eqch70}
\end{align}

To find all NNMs, we can try \textit{all subgroups} of the symmetry group ${G_0=[a,i,u]}$ to choose those displacement patterns, which depend on \textit{only one} arbitrary parameter.

The patterns depending on $m$ arbitrary parameters with $m>1$ form the $m$-dimensional bushes of vibrational modes. Namely in this sense nonlinear normal modes may be called one-dimensional bushes.

In~\cite{DAN1,PhysD98}, three different group-theoretical methods for constructing the bushes of vibrational modes in \textit{arbitrary} $N$-particle nonlinear mechanical systems were developed. The most efficient of these methods uses the concept of irreducible representations of the symmetry groups.

Taking into account the above method that was used for constructing Eq.~(\ref{eqch70}), we conclude that every NNM can be written in the form
\begin{equation}\label{eqch71}
\vec X_{NNM}(t)=\nu(t)\cdot\vec c,
\end{equation}
where $\vec c$ is a certain time-independent vector. Substituting ansatz~(\ref{eqch71}) into the dynamical equations~(\ref{eqch35}--\ref{eqch36}) of the FPU-$\beta$ chain, with explicit forms of the vectors $\vec c$ from Eqs.~(\ref{eqch_60}--\ref{eqch_65}), one can find that FPU-$\beta$ equations are reduced to only one differential equation (governing equation of the corresponding NNM) of the form:
\begin{equation}\label{eqch72}
\ddot\nu+\omega^2\nu+\gamma\frac{\beta}{N}\nu^3=0.
\end{equation}
This is the Duffing equation with different values $\omega$ and $\gamma$ for different NNMs which are listed in Eqs.~(\ref{eqch_60}--\ref{eqch_65}).

Above, we have mentioned the existence of so-called ``dynamical domains'' of all nonlinear normal modes presented in Eqs.~(\ref{eqch_60}--\ref{eqch_65}). Let us comment on this notion borrowed from the theory of phase transitions.

We have already emphasized that a certain symmetry group $G$ corresponds to every NNM. This group is a subgroup of the symmetry group of the considered mechanical system in its equilibrium state ($G\subset G_0$). If we act on the vector $\vec X(t)$ corresponding to a given NNM by operator $g\in G_0$, that \textit{does not} belong to subgroup $G$, we get the \textit{equivalent} configuration vector $\tilde{\vec X}(t)=g\vec X(t)$. The equivalent vector $\tilde{\vec X}(t)$ corresponds to a new NNM, which is described by the \textit{same} dynamical equations as that of the NNM associated with the vector $\vec X(t)$. For example, three dynamical domains are associated with the NNM from Eq.~(\ref{eqch_61}):
\begin{align}
\text{B}[a^3,i]:&\vec c=\frac{3}{\sqrt{6N}}\{1,0,-1~|~1,0,-1~|~1,0,-1~|~\ldots\},\label{eqch90}\\
\text{B}[a^3,ai]:&\vec c=\frac{3}{\sqrt{6N}}\{0,1,-1~|~0,1,-1~|~0,1,-1~|~\ldots\},\label{eqch91}\\
\text{B}[a^3,a^2i]:&\vec c=\frac{3}{\sqrt{6N}}\{1,-1,0~|~1,-1,0~|~1,-1,0~|~\ldots\}.\label{eqch92}
\end{align}

All the displacement patterns (\ref{eqch90}--\ref{eqch92}) differ from each other by a cyclic transposition of the displacements inside each primitive cell of the chain equilibrium state. Let us note that the symmetry groups $G_j$ (${j=1,2,3}$) of NNMs from Eqs.~(\ref{eqch90}--\ref{eqch92}) prove to be \textit{conjugate} subgroups in the parent group $G_0$, for example, $G_2=g^{-1}G_1g$ ($g\in G_0$).

Since the above-discussed ``domains'' possess equivalent dynamical properties, we study below the stability of only one copy of the full set of dynamical domains for every NNM in the FPU-$\beta$ chain.

All symmetry-determined NNMs that can exist in the FPU-$\beta$ chain with an appropriate number of particles are listed in Table~\ref{table10}.

\begin{table*}
\centering
\caption{Nonlinear normal modes in the FPU-$\beta$ chain}\label{table10}
\begin{tabular}{|l|c|c|c|}
  \hline
  NNM          & Displacement      & Modal space                                     & Governing equation \\
               & pattern           & representation                                  & \\
  \hline
  B$[a^2,i]$   & $|1,-1|$          & $\nu\Psi_{N/2}$                                 & $\ddot\nu+4\nu+\frac{16\beta}{N}\nu^3=0$  \\
  B$[a^3,i]$   & $|1,-1,0|$        & $\nu\frac{1}{\sqrt{2}}(\Psi_{N/3}-\Psi_{2N/3})$ & $\ddot\nu+3\nu+\frac{27\beta}{2N}\nu^3=0$ \\
  B$[a^3,iu]$  & $|1,1,-2|$        & $\nu\frac{1}{\sqrt{2}}(\Psi_{N/3}+\Psi_{2N/3})$ & $\ddot\nu+3\nu+\frac{27\beta}{2N}\nu^3=0$ \\
  B$[a^4,ai]$  & $|1,0,-1,0|$      & $\nu\frac{1}{\sqrt{2}}(\Psi_{N/4}-\Psi_{3N/4})$ & $\ddot\nu+2\nu+\frac{4\beta}{N}\nu^3=0$   \\
  B$[a^4,iu]$  & $|1,-1,-1,1|$     & $\nu\Psi_{N/4}$                                 & $\ddot\nu+2\nu+\frac{8\beta}{N}\nu^3=0$   \\
  B$[a^3u,ai]$ & $|1,1,0,-1,-1,0|$ & $\nu\frac{1}{\sqrt{2}}(\Psi_{N/6}-\Psi_{5N/6})$ & $\ddot\nu+\nu+\frac{3\beta}{2N}\nu^3=0$   \\
  \hline
\end{tabular}
\end{table*}

\section{Stability diagrams for NNMs in the FPU-$\beta$ chain}

The detailed numerical analysis of stability of NNMs in the FPU chains can be found in~\cite{FPU2}, where results have been presented in the form of certain ``stability diagrams''. Let us consider the structure of such diagrams and the method of their obtaining.

Every NNM represents a \textit{periodic} dynamical regime and the standard Floquet method can be applied for investigation of its stability. According to this method, we have to linearize the nonlinear FPU-$\beta$ dynamical equations~(\ref{eqch35},\ref{eqch36}) in the vicinity of a given NNM $\vec X_{NNM}(t)$. To this end, let us introduce an infinitesimal vector
\begin{equation}\label{eqch79}
\vec\delta=\{\delta_1,\delta_2,\delta_3,\ldots,\delta_N\},
\end{equation}
that determines a certain \textit{perturbation} of the NNM. Letting
\begin{equation}\label{eqch80}
\vec X(t)=\vec X_{NNM}(t)+\vec\delta(t),
\end{equation}
we substitute $\vec X(t)$ into nonlinear equations~(\ref{eqch35},\ref{eqch36}) and omit all nonlinear terms in $\delta_j(t)$ (${j=1..N}$). As a result of this procedure, we get the linearized (variational) system
\begin{equation}\label{eqch81}
\ddot{\vec\delta}=J(t)\,\vec\delta,
\end{equation}
where $J(t)$ is the corresponding Jacobian matrix (see details in~\cite{FPU2} and especially in~\cite{Zhukov}).

Equation~(\ref{eqch81}) represents an ${N\times N}$ system of linear differential equations with time-dependent coefficients, which, in turn, are determined by the periodic function $\nu(t)$ describing the time-evolution of the considered NNM.

\begin{figure*}
\centering
\resizebox{\textwidth}{!}{
\setlength{\unitlength}{0.6mm}
\baselineskip 30mm
\begin{tabular}{rcc}
a)
&
\begin{picture}(130,50)(0,0)
\put(5,5){\includegraphics[width=120\unitlength,height=36\unitlength]{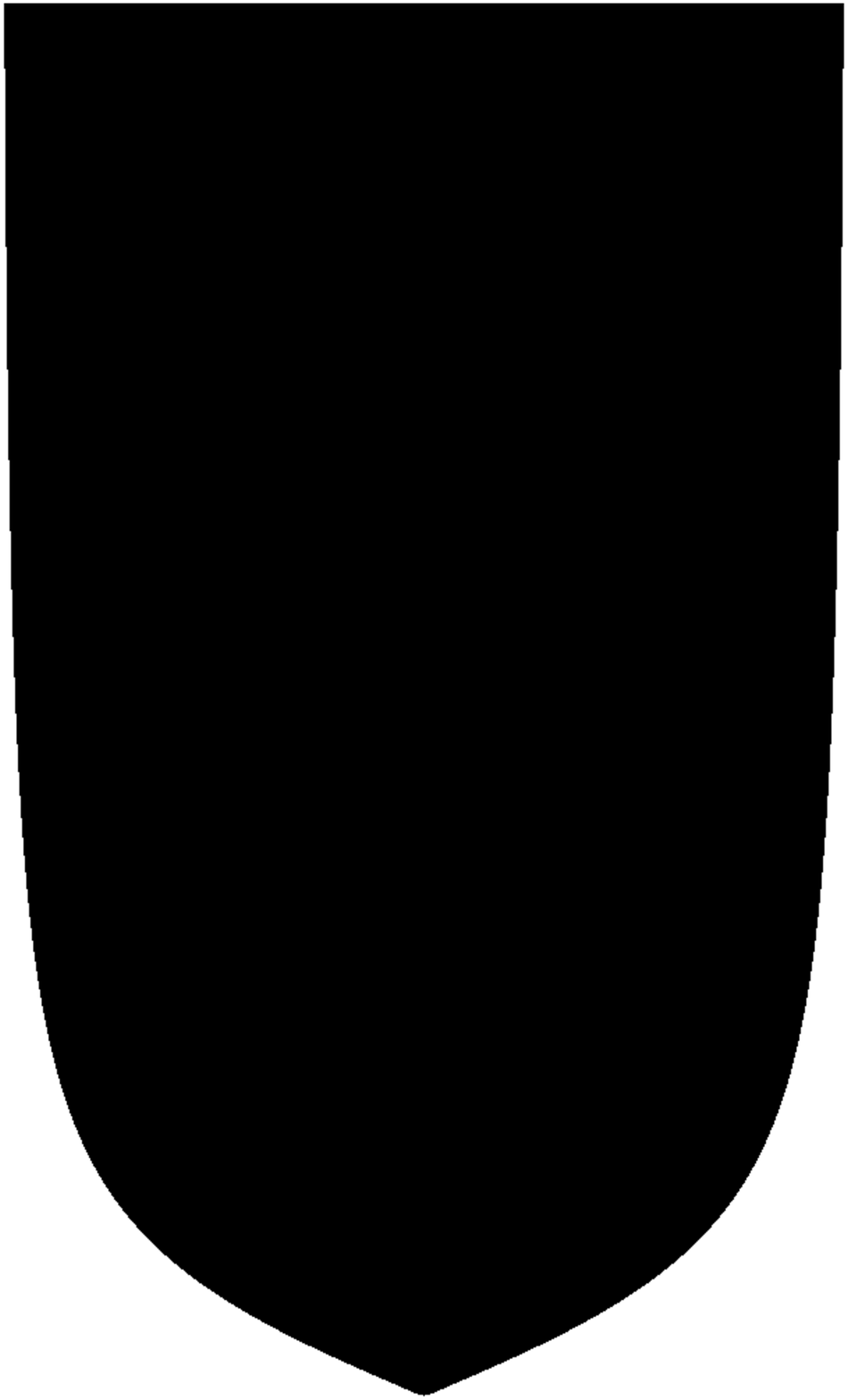}}
\drawgrid
\put(65,5){\line(0,1){36}}
\end{picture}
&
\begin{picture}(130,50)(0,0)
\put(5,5){\includegraphics[width=120\unitlength,height=36\unitlength]{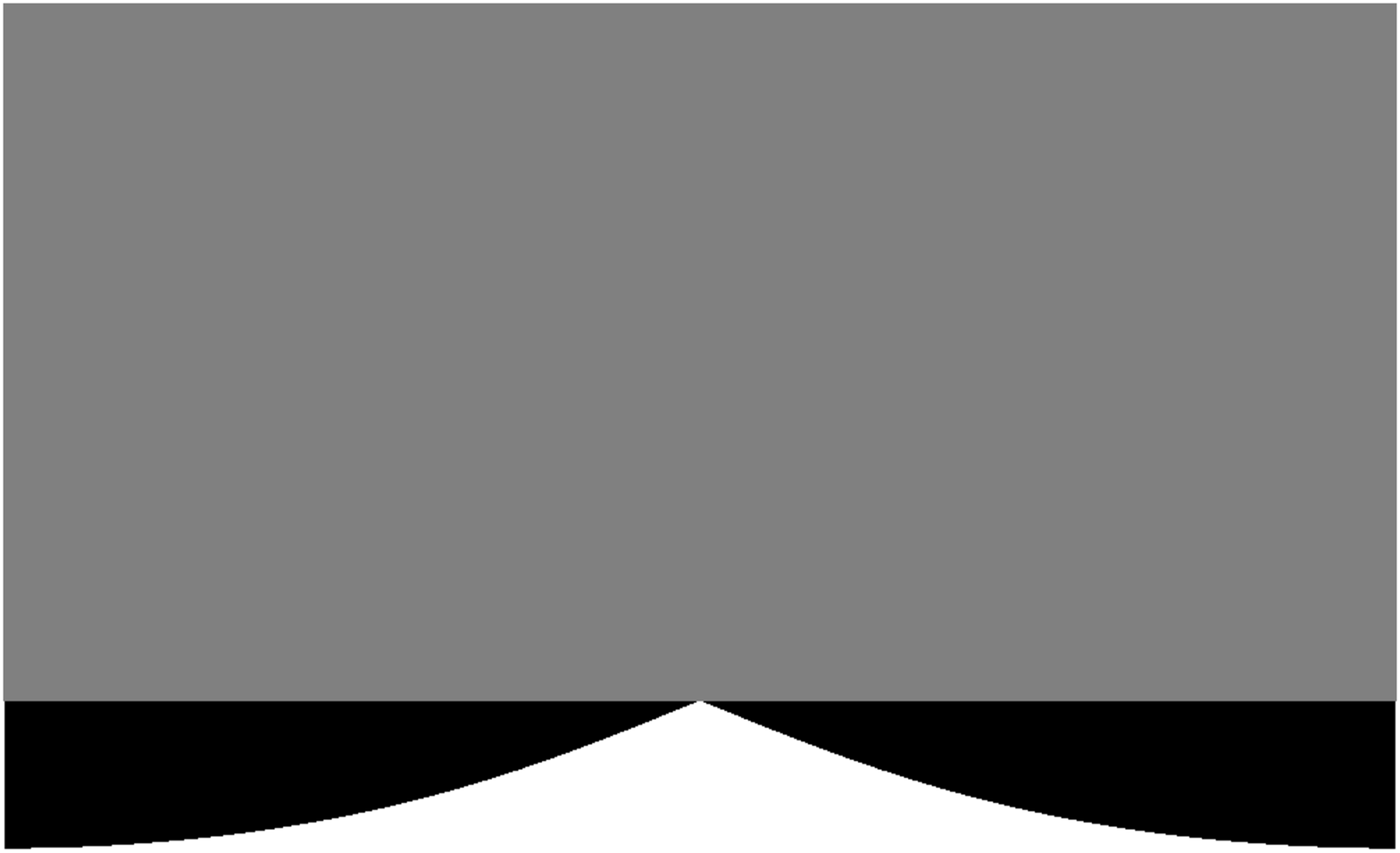}}
\drawgridA
\put(65,5){\line(0,1){36}}
\end{picture}
\\
b)
&
\begin{picture}(130,50)(0,0)
\put(5,5){\includegraphics[width=120\unitlength,height=36\unitlength]{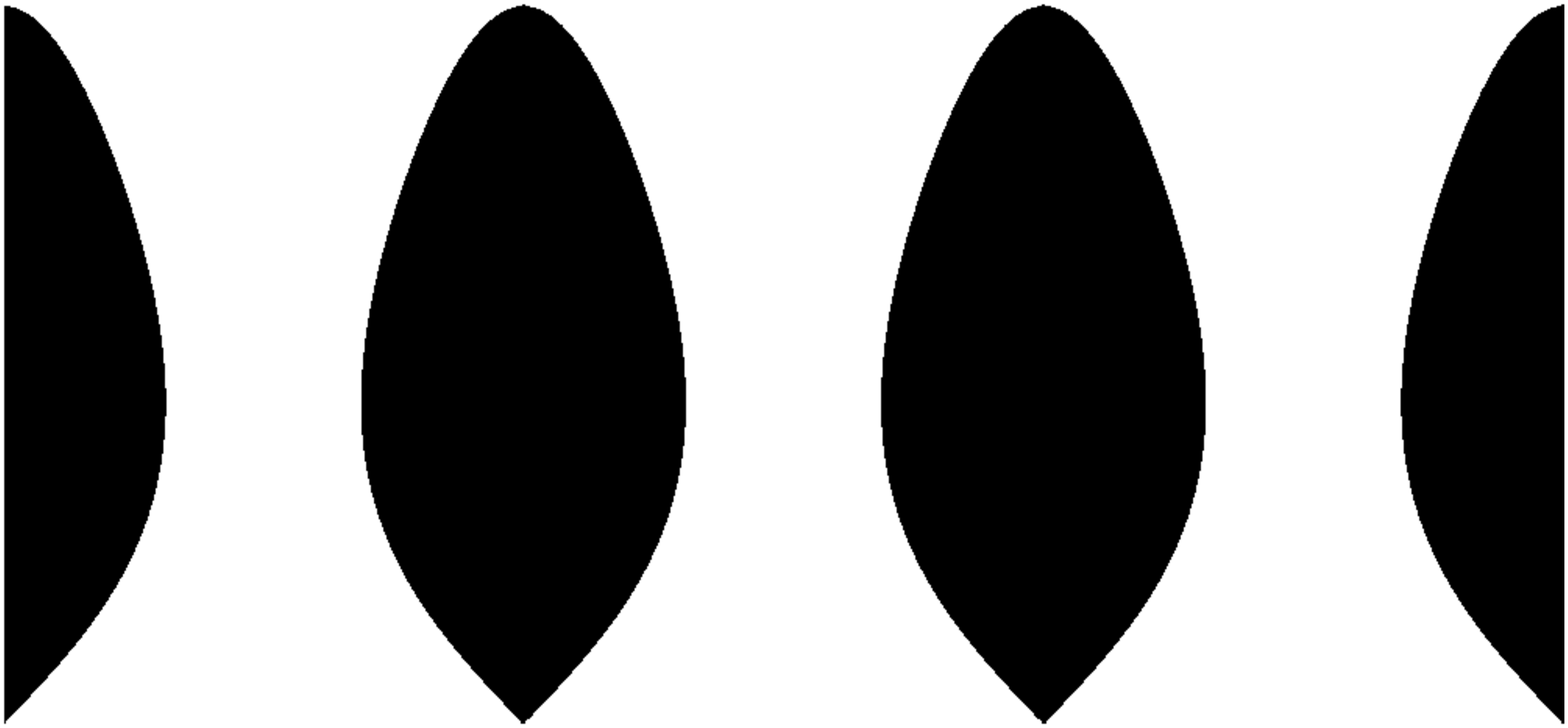}}
\drawgridB
\put(45,5){\line(0,1){36}}
\put(85,5){\line(0,1){36}}
\end{picture}
&
\begin{picture}(130,50)(0,0)
\put(5,5){\includegraphics[width=120\unitlength,height=36\unitlength]{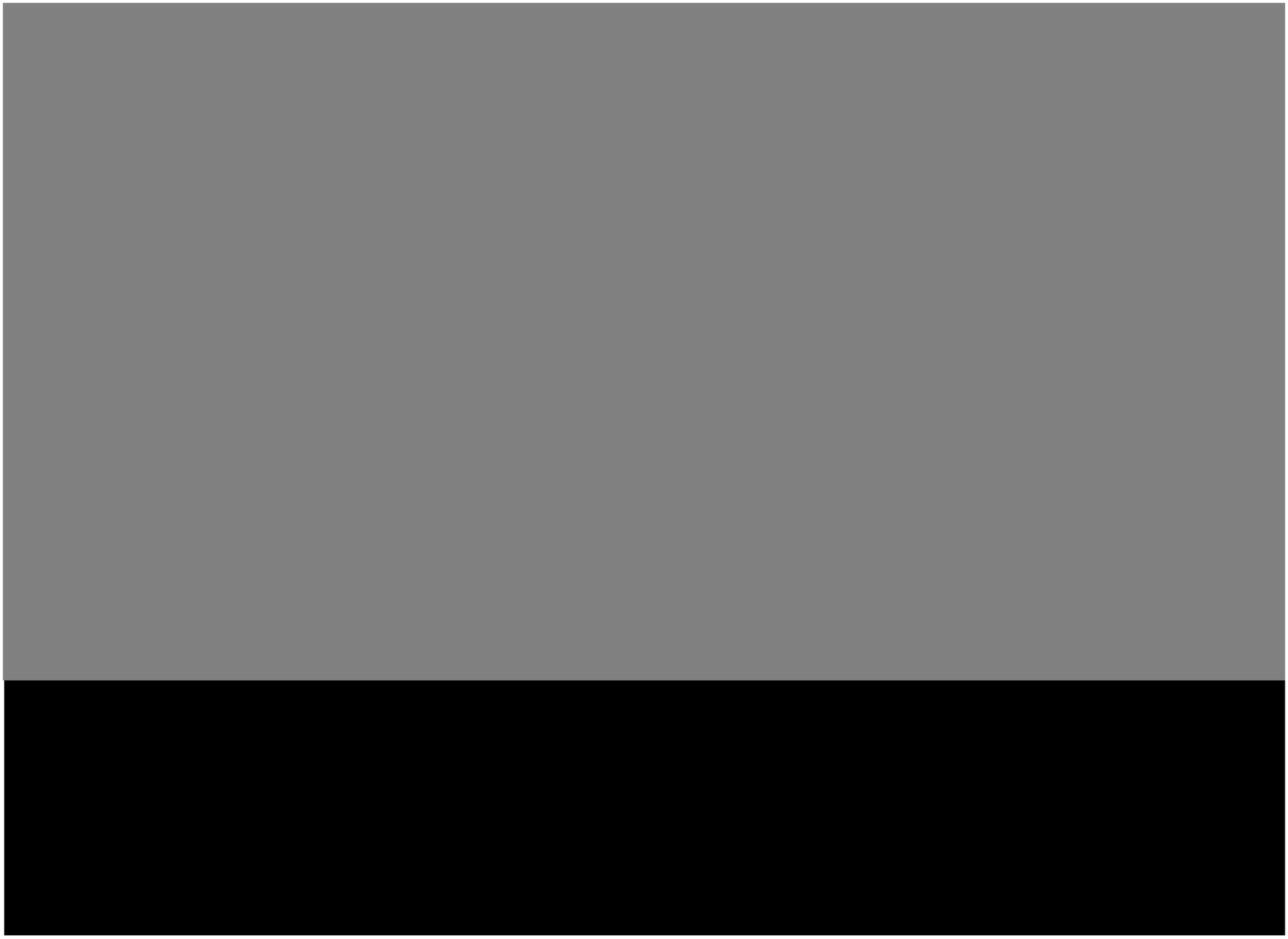}}
\drawgridA
\put(45,5){\line(0,1){36}}
\put(85,5){\line(0,1){36}}
\end{picture}
\\
c)
&
\begin{picture}(130,50)(0,0)
\put(5,5){\includegraphics[width=120\unitlength,height=36\unitlength]{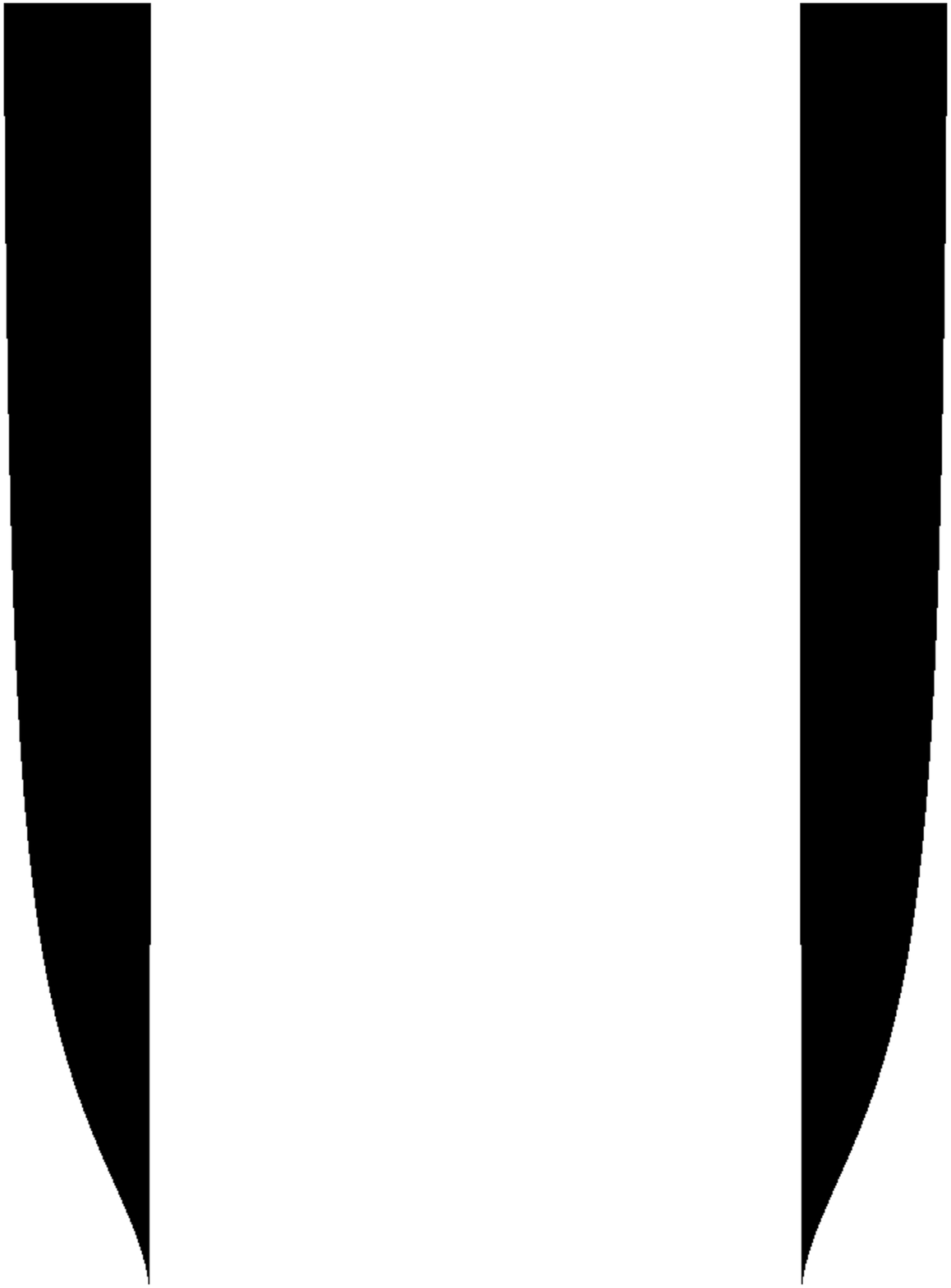}}
\drawgrid
\put(35,5){\line(0,1){36}}
\put(95,5){\line(0,1){36}}
\end{picture}
&
\begin{picture}(130,50)(0,0)
\put(5,5){\includegraphics[width=120\unitlength,height=36\unitlength]{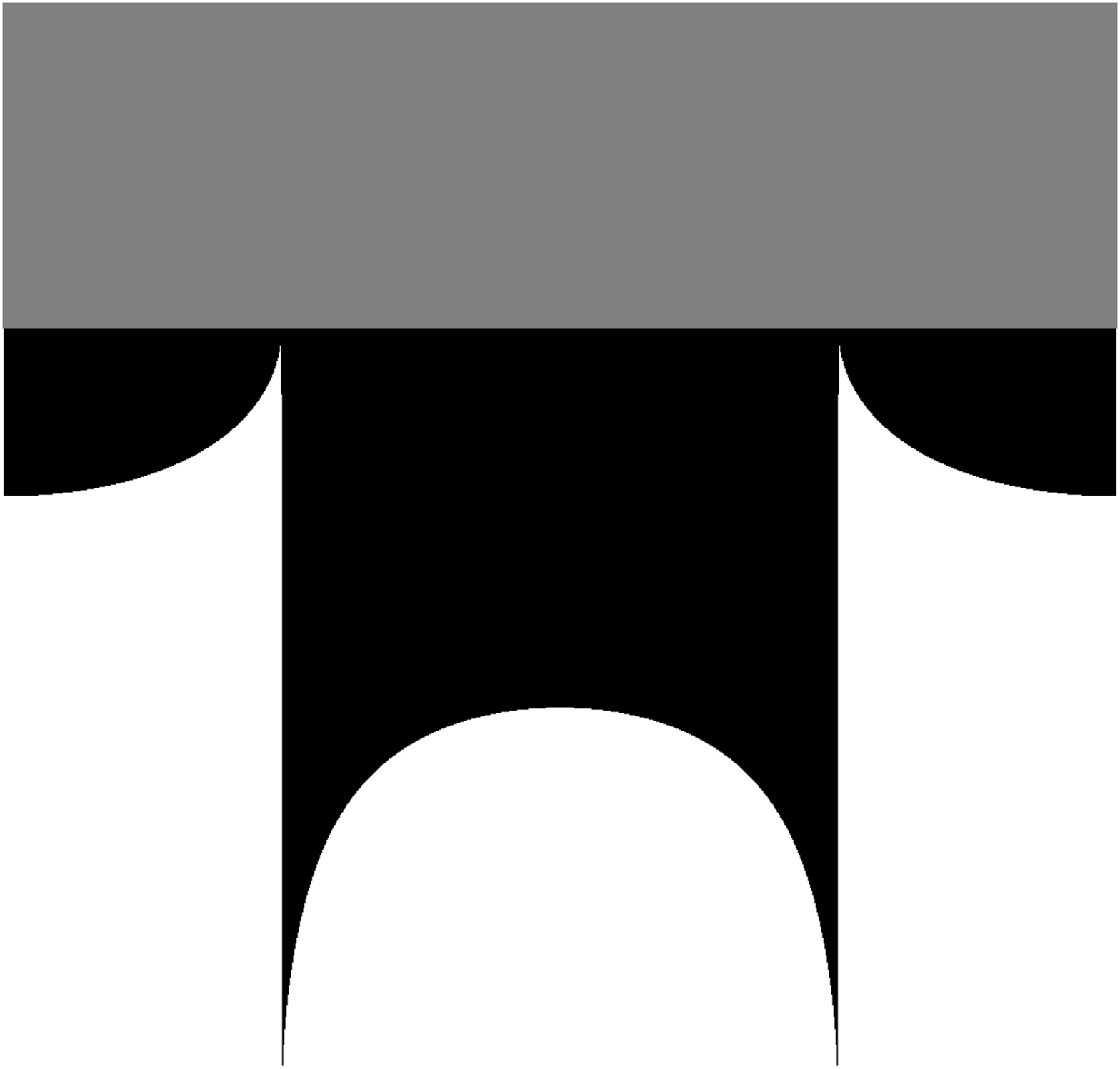}}
\drawgridA
\put(35,5){\line(0,1){36}}
\put(95,5){\line(0,1){36}}
\end{picture}
\\
d)
&
\begin{picture}(130,50)(0,0)
\put(5,5){\includegraphics[width=120\unitlength,height=36\unitlength]{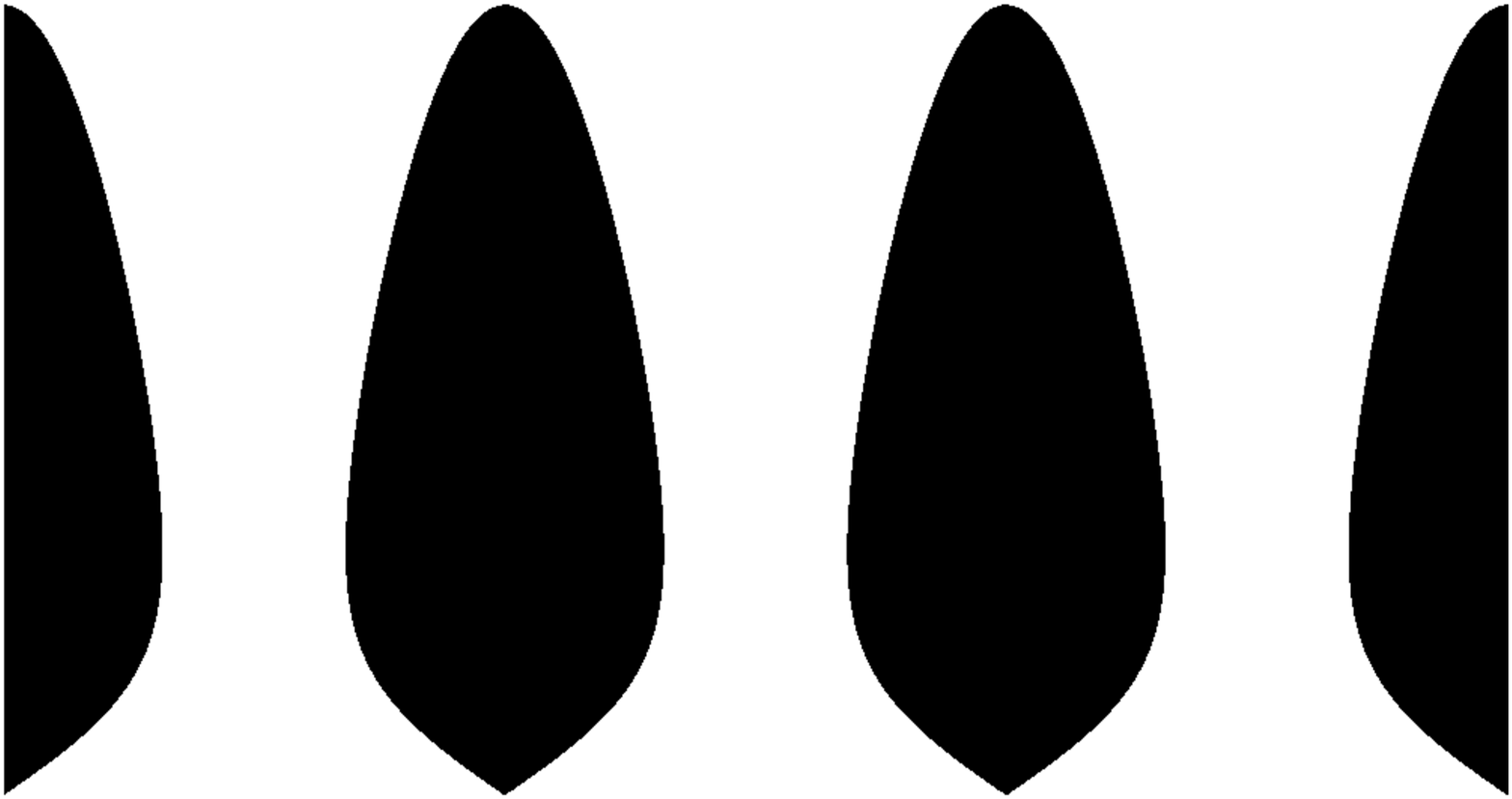}}
\drawgrid
\put(45,5){\line(0,1){36}}
\put(85,5){\line(0,1){36}}
\end{picture}
&
\begin{picture}(130,50)(0,0)
\put(5,5){\includegraphics[width=120\unitlength,height=36\unitlength]{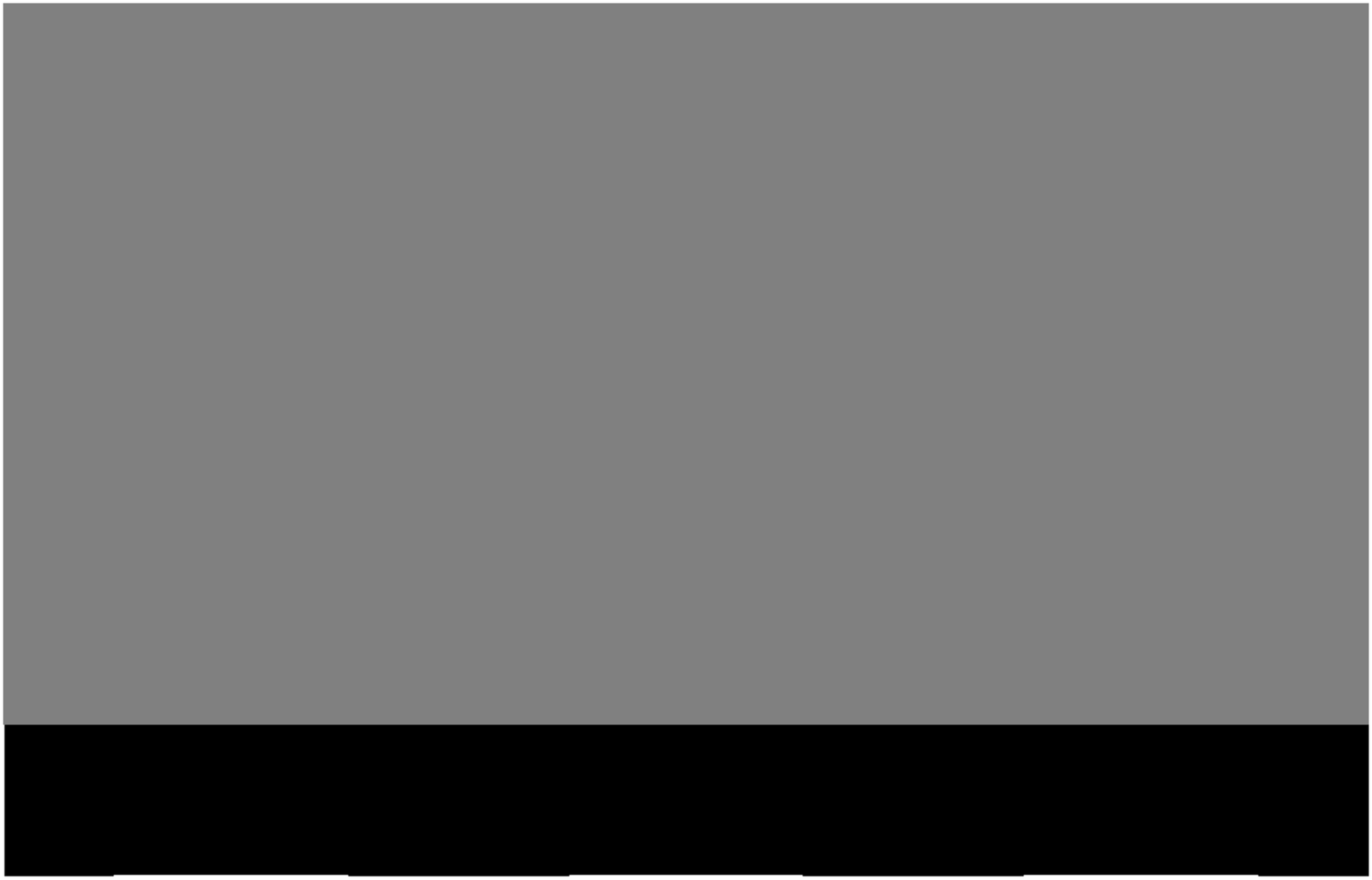}}
\drawgridA
\put(45,5){\line(0,1){36}}
\put(85,5){\line(0,1){36}}
\end{picture}
\\
e)
&
\begin{picture}(130,50)(0,0)
\put(5,5){\includegraphics[width=120\unitlength,height=36\unitlength]{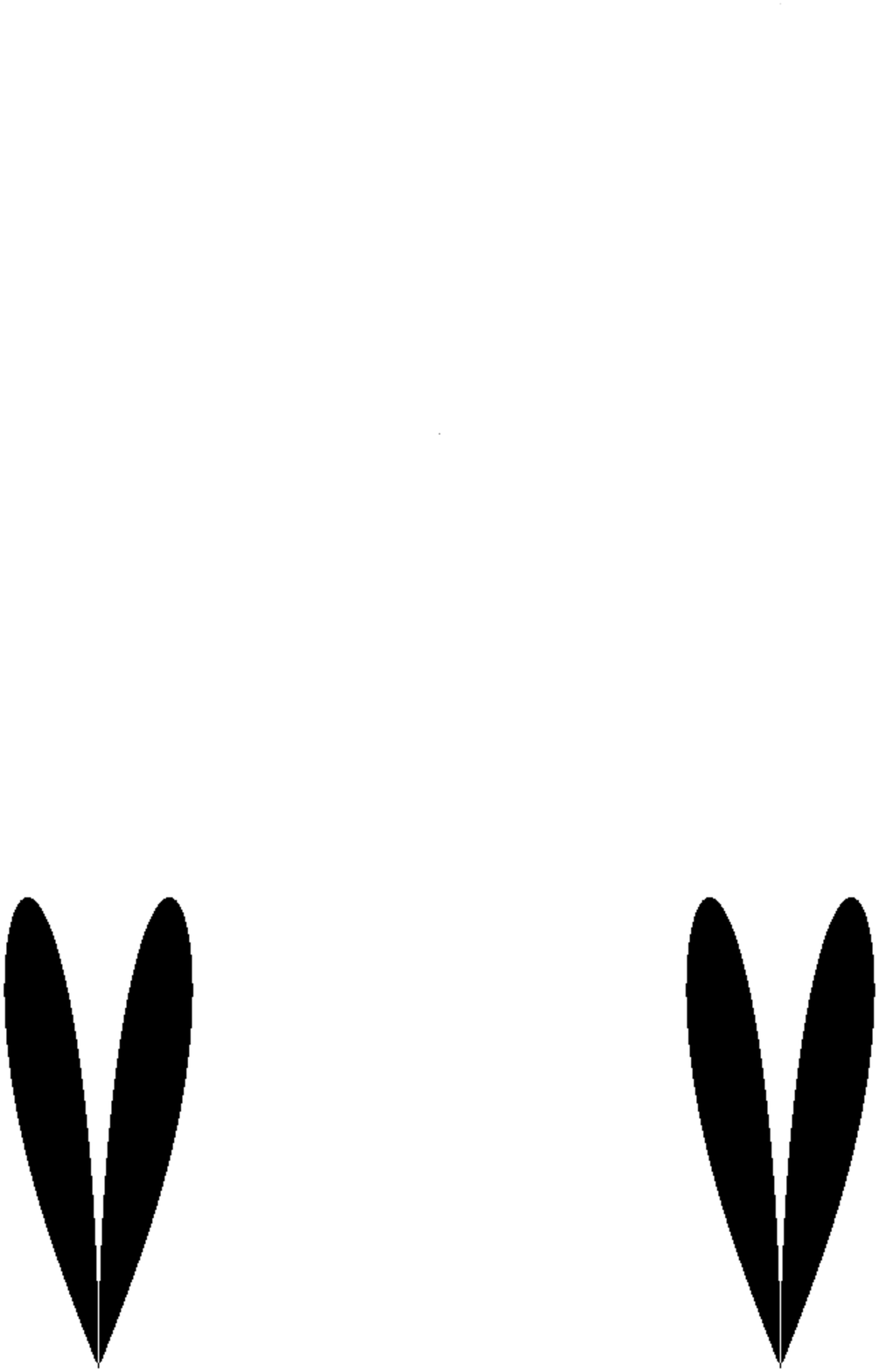}}
\drawgridA
\put(35,5){\line(0,1){36}}
\put(95,5){\line(0,1){36}}
\end{picture}
&
\begin{picture}(130,50)(0,0)
\put(5,5){\includegraphics[width=120\unitlength,height=36\unitlength]{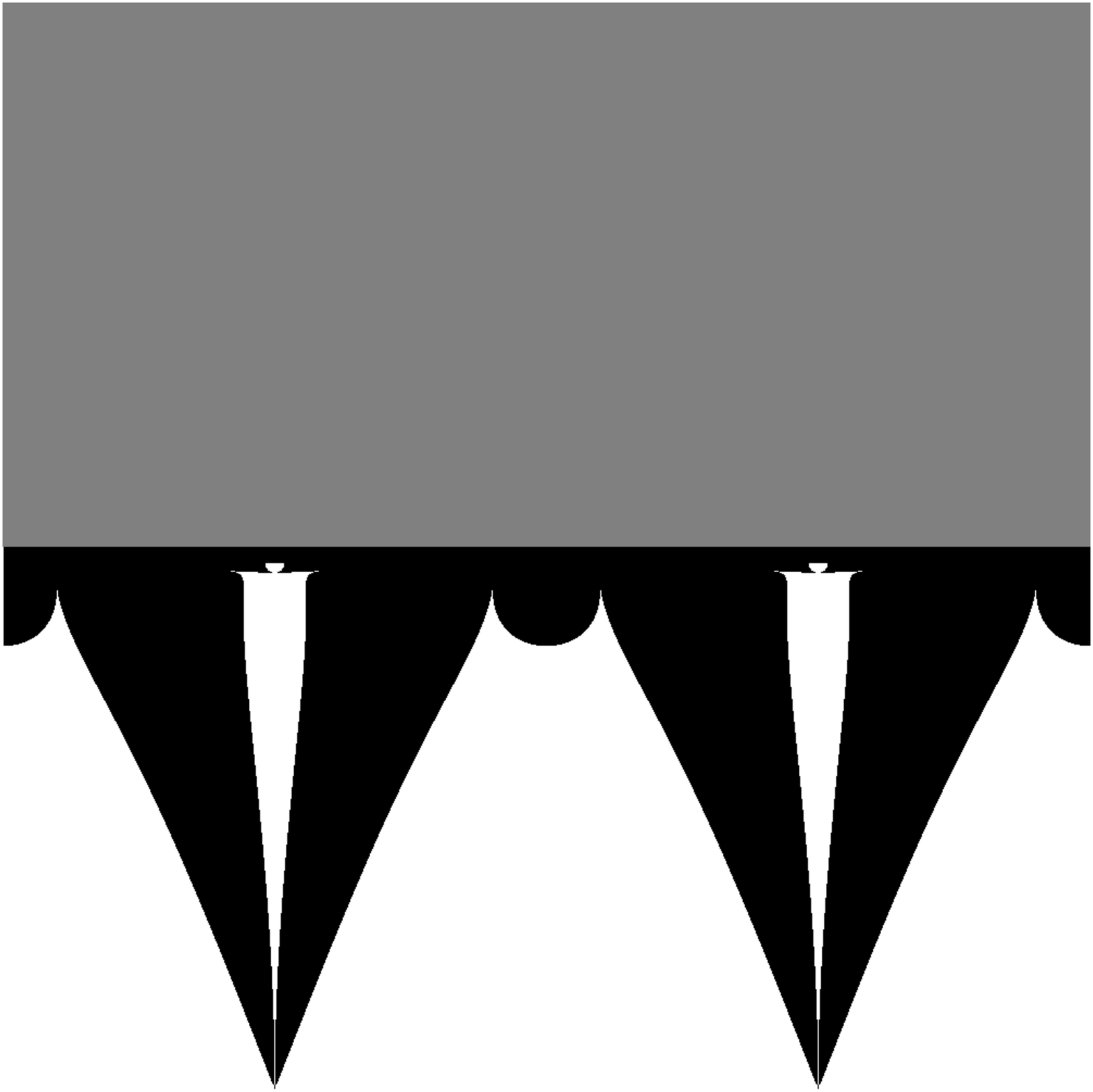}}
\drawgridA
\put(35,5){\line(0,1){36}}
\put(95,5){\line(0,1){36}}
\end{picture}
\\
f)
&
\begin{picture}(130,50)(0,0)
\put(5,5){\includegraphics[width=120\unitlength,height=36\unitlength]{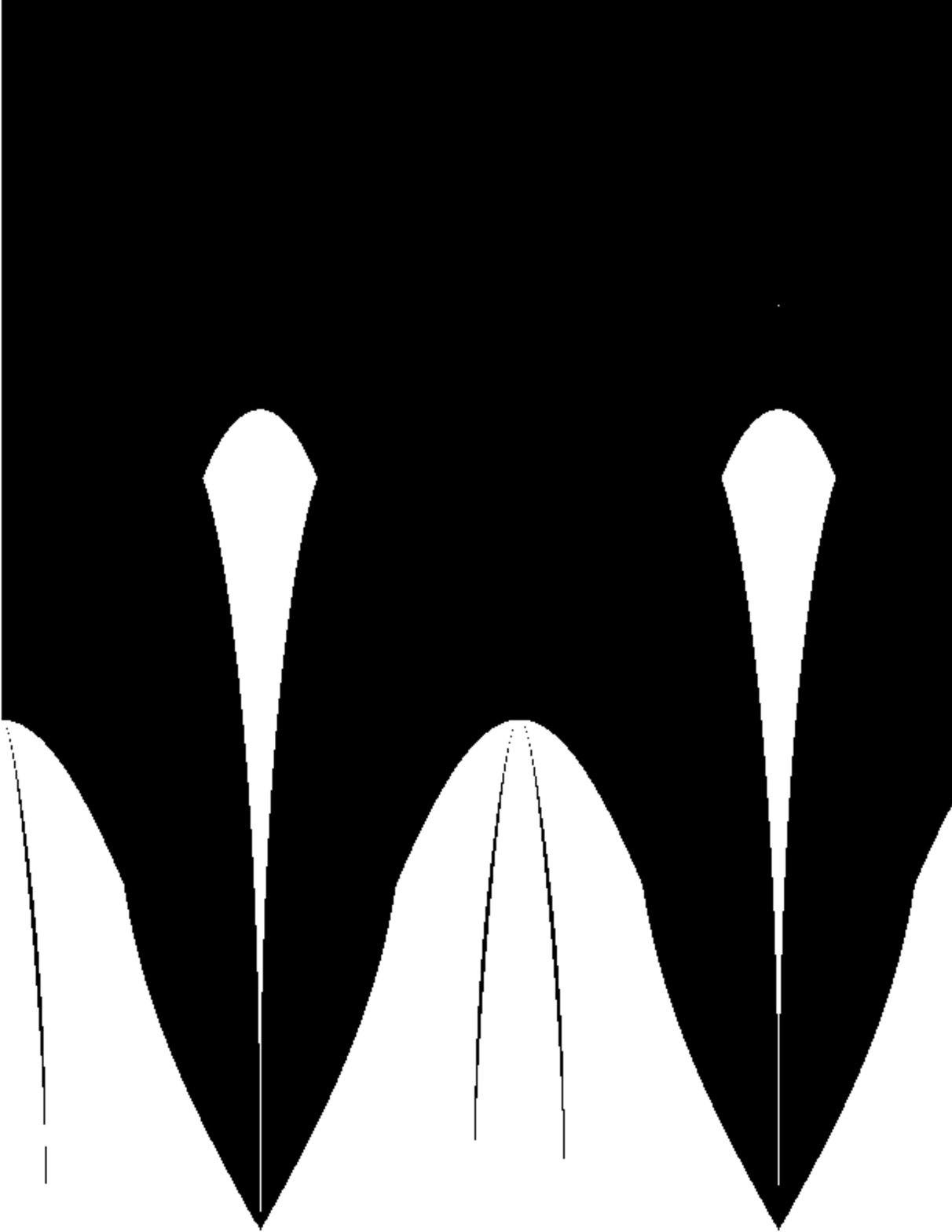}}
\drawgridB
\put( 25,5){\line(0,1){36}}
\put(105,5){\line(0,1){36}}
\end{picture}
&
\begin{picture}(130,50)(0,0)
\put(5,5){\includegraphics[width=120\unitlength,height=36\unitlength]{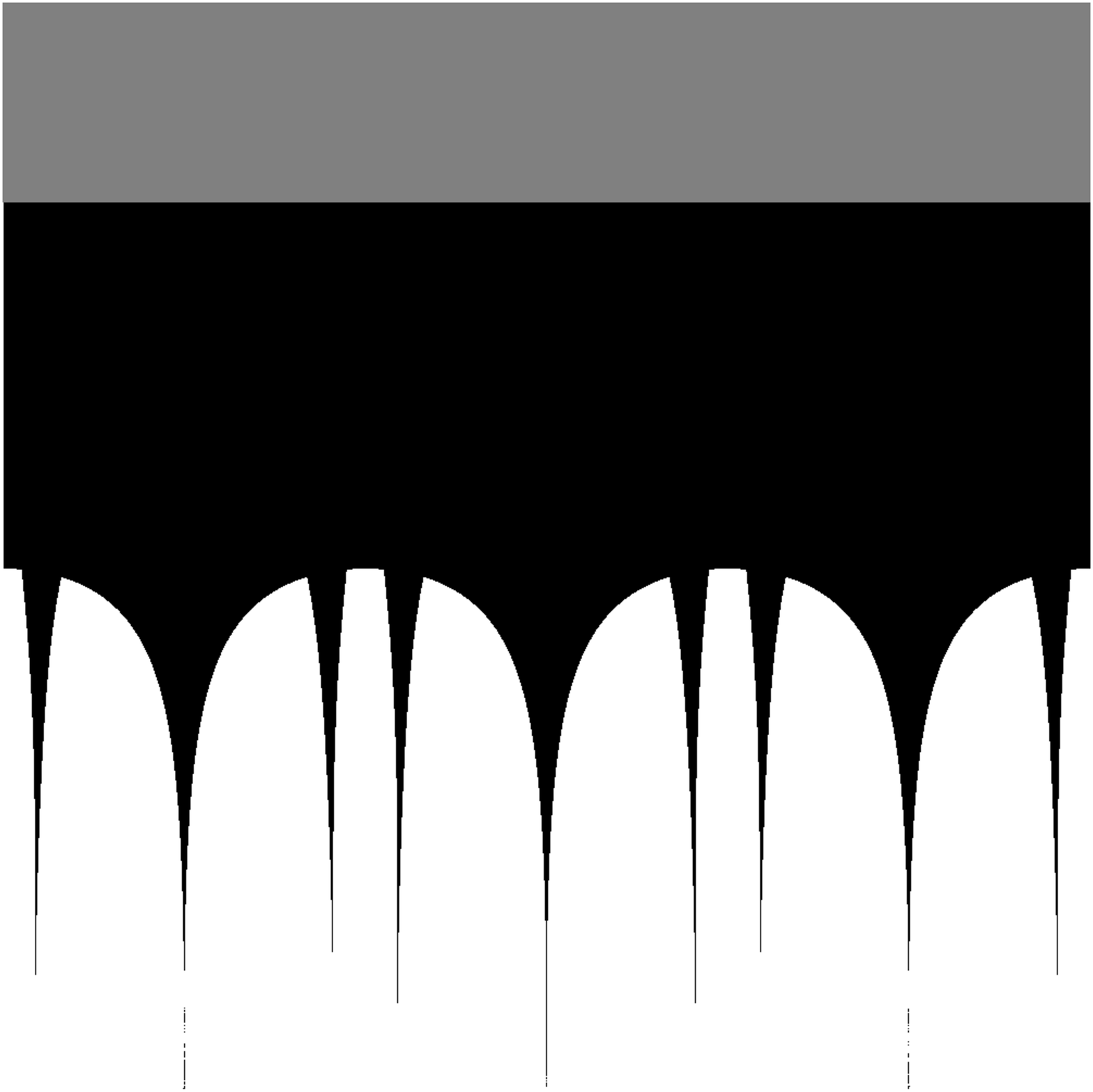}}
\drawgridA
\put( 25,5){\line(0,1){36}}
\put(105,5){\line(0,1){36}}
\end{picture}
\end{tabular}
}

\caption{Stability diagrams for NNMs in the FPU-$\beta$ chain (left $\beta>0$, right $\beta<0$): a)~B$[a^2,i]$; b)~B$[a^3,i]$; c)~B$[a^4,ai]$; d)~B$[a^3,iu]$; e)~B$[a^4,iu]$; f)~B$[a^3u,ai]$.\label{figStabDiag}}
\end{figure*}

In~\cite{Zhukov}, we have developed a group-theoretical method for \textit{splitting} the linearized system ${\ddot{\vec\delta}=J(t)\,\vec\delta}$ into a number of independent subsystems of small dimensions (for all NNMs in the FPU-$\beta$ chain these dimensions do not exceed~3). In the explicit form, this splitting can be found for all NNMs in the FPU-$\beta$ chain in Table~8 of the paper~\cite{FPU2}. Let us comment on that table using as an example the nonlinear normal mode B$[a^4,iu]$ for $\beta>0$. The ``rabbit ears'' stability diagram, depicted in Fig.~\ref{figStabDiag}, corresponds to this mode.

In this case, the linearized system can be splitted into $\left(\frac{N}{2}-2\right)$ two-dimensional systems of the following form
\begin{widetext}
\begin{equation}\label{eqch82}
\begin{array}{rcl}
\displaystyle\ddot\delta_j+4\sin^2\left(\frac{\pi j}{N}\right)\left[1+\frac{6\beta}{N}\nu^2(t)\right]\delta_j &=& \displaystyle-\frac{12\beta}{N}\nu^2(t)\sin\left(\frac{2\pi j}{N}\right)\delta_{j'},\\[8pt]
\displaystyle\ddot\delta_{j'}+4\sin^2\left(\frac{\pi j'}{N}\right)\left[1+\frac{6\beta}{N}\nu^2(t)\right]\delta_{j'} &=& \displaystyle-\frac{12\beta}{N}\nu^2(t)\sin\left(\frac{2\pi j}{N}\right)\delta_j.
\end{array}
\end{equation}
\end{widetext}
Here $j'=\frac{N}{2}-j$, and ${j=1,\ldots,\left(\frac{N}{2}-1\right)}$.

Thus, the $j$th component of the infinitesimal vector $\vec\delta$ (after the appropriate orthogonal transformation) turns out to be coupled only with its $j'$~component, where ${j'=\frac{N}{2}-j}$. Therefore, the investigation of stability of the nonlinear normal mode B$[a^4,iu]$ reduces to studying stability of \textit{zero solution} of Eqs.~(\ref{eqch82}) with different values of the indices $j$ and ${j'=\frac{N}{2}-j}$. It can be seen from Eqs.~(\ref{eqch82}) that the pairs of modes whose indices are situated symmetrically with respect to the index $j_0=\frac{N}{4}$ of the considered NNM are coupled.

Note that Eqs.~(\ref{eqch82}) represent the system of two coupled differential equations with time-periodic coefficients determined by the solution $\nu(t)$ of the Duffing equation~(\ref{eqch72}). It is well known that equations of such a type possess domains of stable and unstable motion.\footnote{In~\cite{FPU2}, we have illustrated this idea with the case of full splitting for the NNM B$[a^2,i]$ in the FPU-$\alpha$ chain. In this case, the linearized system $\ddot{\vec\delta}=J(t)\,\vec\delta$ can be decomposed into individual Mathieu equations for which the stability diagram is well known.}

For the system~(\ref{eqch82}), we can construct the stability diagram in the plane (${j-A}$), where $A$ is the amplitude of the NNM [i.e.\ the amplitude of the function $\nu(t)$ from Eq.~(\ref{eqch72})], while $j$ is its index. Such diagrams provide us with the most valuable information for the case of the ``thermodynamic limit'' when the number of lattice cells tends to infinity (${N\rightarrow\infty}$).

When ${N\rightarrow\infty}$, the number of modes also tends to infinity and it is convenient to introduce the wave number $k$ (${0\le k<2\pi}$) instead of the mode index $j$:
\begin{equation}\label{eqch190}
k=\left(\frac{2\pi}{N}\right)j,\quad j=1..N.
\end{equation}

In Eq.~(\ref{eqch10}) there is a normalizing coefficient $1/\sqrt{N}$ originating from the conventional definition of the mode norm $|\vec\Psi_j|^2=1$. As a consequence, displacements of all particles corresponding to a given normal mode with fixed amplitude tend to zero with $N\to\nobreak\infty$, while the coefficient $1/N$ appears in Eqs.~(\ref{eqch72}), (\ref{eqch82}) etc.\ in combination with the coefficient~$\beta$.

Hereafter, we use a more convenient for our purposes normalization of normal modes (normalization to the ``volume'' of the system), namely, we assume $|\vec\Psi_j|^2=N$. This permits us to exclude the coefficient $1/N$ coupled to $\beta$. Moreover, using a trivial coordinate rescaling, one can set $|\beta|=1$.

As a result, Eqs.~(\ref{eqch82}) are transformed to the following form:
\begin{equation}\label{eqch82free}
\begin{array}{rcl}
\ddot\delta_k+4\sin^2(\frac{k}{2})[1\pm 6\nu^2(t)]\delta_k&=&\mp 12\nu^2(t)\sin(k)\,\delta_{k'},\\
\ddot\delta_{k'}+4\cos^2(\frac{k}{2})[1\pm 6\nu^2(t)]\delta_{k'}&=&\mp 12\nu^2(t)\sin(k)\,\delta_k.
\end{array}
\end{equation}
Here the upper and lower signs correspond to $\beta>0$ and $\beta<0$, respectively.

Following this idea, we construct our stability diagrams in the (${k-A}$) plane, where $A$ is the amplitude of a normal mode subjected to the above mentioned normalizing condition $|\vec\Psi_j|^2=N$. For any fixed $N$, the permissible values of the wave number $k$ represent the equidistant set inside the interval $[0,2\pi]$. When ${N\rightarrow\infty}$, these values of $k$ form a \textit{dense} set on the above interval, depicted on the horizontal axis of diagrams in Fig.~\ref{figStabDiag}.

Rich information about stability properties of NNMs can be obtained from diagrams of such a type. Let us consider this idea in more detail using as an example the stability diagram of the NNM B$[a^4,iu]$ depicted in Fig.~\ref{figStabDiag}e for $\beta>0$.

In this diagram, we denote the permissible values of the wave number $k$ for ${N=12}$ by the dotted vertical lines (${k=\gamma,}\,2\gamma,\,3\gamma,\,\ldots,\,12\gamma$, where ${\gamma=2\pi/12}$). The black color corresponds to the regions of unstable motion in the plane (${k-A}$). From Fig.~\ref{figStabDiag}e one can see that for the FPU-$\beta$ chain with ${N=12}$ particles the NNM B$[a^4,iu]$ with ${\vec c=\{1,-1,-1,1~|~1,-1,-1,1~|~1,-1,-1,1\}}$ turns out to be \textit{stable} for \textit{every} amplitude $A$ up to 2 (actually, at least, up to ${A=20}$). Indeed, the dotted vertical lines do not cross the black (unstable) regions in the form of the rabbit ears.

Obviously, the mode B$[a^4,iu]$ is also stable for the chains with ${N=4}$ and ${N=8}$ particles since the vertical dotted lines, similar to those depicted in Fig.~\ref{figStabDiag}e, are \textit{more distant} from each other than for ${N=12}$, and the regions of instability situate fully between the neighboring lines.

In the case ${N=16}$, there exist such dotted vertical lines that are \textit{tangents} to the rabbit-ears unstable regions, while for $N>16$ (note that the relation ${N\bmod 4=0}$ must hold!) these lines begin to \textit{cross} the unstable regions. Therefore, we can conclude that the considered nonlinear normal mode in the FPU-$\beta$ chains with ${N>16}$ particles becomes unstable for the vibrational amplitudes, which fall into the black regions in Fig.~\ref{figStabDiag}e. The case ${N=16}$ represents the ``boundary'' case between stable and unstable behavior of the B$[a^4,iu]$ NNM for appropriate amplitudes.

From Fig.~\ref{figStabDiag}e, it can be seen that the \textit{critical} amplitude $A_c$ of the nonlinear normal mode B$[a^4,iu]$, for which this mode loses its stability, decreases with increasing $N$ and we can evaluate numerically the corresponding scaling law of the function $A_c(N)$ in the thermodynamic limit $N\rightarrow\infty$.

Let us consider this question in more detail. The instability regions touch the $k$-axis at the points $\frac{\pi}{2}$ $\left(j=\frac{N}{4}\right)$ and $\frac{3\pi}{2}$ $\left(j=\frac{3N}{4}\right)$, which correspond to the NNM B$[a^4,iu]$ and its dynamical domain B$[a^4,a^2iu]$, respectively. Namely near these points the considered NNMs loses their stability in the thermodynamic limit $N\rightarrow\infty$.

Let us focus on the point ${\left(k=\frac{\pi}{2}\right)}$, which corresponds to the index $j_0=\frac{N}{4}$ of the mode B$[a^4,iu]$. All the neighboring $k$-points correspond to the ``sleeping'' modes, i.e.\ modes with zero amplitudes for the stable behavior of the mode B$[a^4,iu]$. If we increase the amplitude $A$ of the NNM B$[a^4,iu]$ from zero, the sleeping modes, which become firstly excited, are those with indices ${j'=\frac{N}{4}\pm 1}$, i.e.\ the modes closest in number to the index ${j_0=\frac{N}{4}}$ of the NNM B$[a^4,iu]$.

We denote by $A_c(N)$ the critical value of the amplitude $A$ of the considered nonlinear normal mode for which the loss of its stability takes place, i.e.\ $A_c(N)$ is the threshold of the B$[a^4,iu]$ NNM stability for a given~$N$. Note that we can speak about the threshold value $A_c^{(j)}$ for the \textit{excitation} of the sleeping $j$-mode as a result of its interaction with the B$[a^4,iu]$ NNM whose index is equal to ${j_0=\frac{N}{4}}$. Then from Fig.~\ref{figStabDiag}e, we obtain
\begin{equation}\label{eqch77}
A_c(N)=\min_j A_c^{(j)}=A_c^{(\frac{N}{4}\pm 1)}.
\end{equation}

For the index $j$ situated near ${j_0=\frac{N}{4}}$ $\left(k=\frac{\pi}{2}\right)$, one can see from Fig.~\ref{figStabDiag}e, that
\begin{equation}\label{eqch78}
A_c^{(j)}=\Delta k\cdot\tan\alpha,
\end{equation}
where ${\Delta k=\frac{2\pi}{N}}$ is the minimal distance between the neighboring permitted points on the $k$-axis and $\tan\alpha$ is determined by the tangent to the black unstable region near the point ${k_0=\frac{\pi}{2}}$.

The scaling law of the function $A_c(N)$ for $N\rightarrow\infty$ can be deduced from Eqs.~(\ref{eqch77}), (\ref{eqch78}):
\begin{equation}\label{eqch79b}
A_c(N)=\frac{2\pi}{N}\tan\alpha\quad(N\rightarrow\infty).
\end{equation}
Here ${\tan\alpha\sim 1}$, as it might be evaluated from Fig.~\ref{figStabDiag}e by the straightforward manner.

The above discussed threshold function $A_c(N)$ for $N\rightarrow\infty$ can be calculated analytically. For the nonlinear normal mode B$[a^4,iu]$, such calculation has been recently performed in~\cite{LeoLeo2007}. The similar analytical results for the $\pi$-mode $\left(j_0=\frac{N}{2}\right)$ were earlier obtained in~\cite{Budinsky1983,SanduskyPage1994,Flach1996,PoggiRuffo1997,Dauxois1997,Yoshimura2004,Dauxois2005}.

In the present paper, we present the \textit{analytical} estimations of stability thresholds for ${N\rightarrow\infty}$ for \textit{all possible} in the FPU-$\beta$ (${\beta>0}$ and ${\beta<0}$) nonlinear normal modes (see Sec.~5).

Let us return to the stability diagram in Fig.~\ref{figStabDiag}e for the NNM B$[a^4,iu]$ for $\beta>0$. It is very interesting that for the \textit{sufficiently large} amplitudes (${A>A_c\approx 0.346}$) the nonlinear normal mode B$[a^4,iu]$ again becomes stable \textit{even in the thermodynamic limit} ${N\rightarrow\infty}\,$! Results of such a kind cannot be obtained by the analytical methods applied in~\cite{LeoLeo2007} and in the present paper. However, the stability of NNMs for nonlinear chains in the case of large amplitudes can be analyzed by the numerical method used in~\cite{FPU2} which provide us the stability diagrams depicted in Fig.~\ref{figStabDiag}.

In the recent paper~\cite{LeoLeo2011}, the existence of the ``second stability threshold'' $E_c$ for the NNM B$[a^4,iu]$ was also revealed. Beyond $E_c$, the stability of the mode B$[a^4,iu]$ (in~\cite{LeoLeo2011}, it is called $\pi/2$-mode) is restored. This result was obtained by numeric methods. In term of the partial energy per one particle of the FPU-$\beta$ chain, this threshold turns out to be equal to $E_c=0.14715$. On the other hand, we can find an approximate value for $E_c$ directly from diagram depicted in Fig.~\ref{figStabDiag}e: $E_c\approx 0.15$. This value is in good agreement with that from~\cite{LeoLeo2011}.

Moreover, such a ``surprising behaviour'' (as has been written in~\cite{LeoLeo2011}) one can reveal directly from our stability diagrams presented in~\cite{FPU2} for NNMs B$[a^3,i]$ and B$[a^3,iu]$. We reproduce these diagrams in Fig.~\ref{figStabDiag}b and \ref{figStabDiag}d of the present paper, and one can found from them the approximate values of the ``second stability thresholds'' for the above mentioned nonlinear normal modes: $E_c($B$[a^3,i])\approx 3.70$ and $E_c($B$[a^3,iu])\approx 24.8$.

In the right column of Fig.~\ref{figStabDiag}, we present new results on the stability of all possible NNMs in the FPU-$\beta$ chain with $\beta<0$. In contrast to the case $\beta>0$, the chain with $\beta<0$ occurs to be unstable itself for the mode amplitudes which exceed a certain critical value (this case corresponds to the region depicted by gray color in Fig.~\ref{figStabDiag}). Note that beyond $E=0.092$ for the NNM B$[a^3,i]$ and $E=0.14$ for the NNM B$[a^3,iu]$, all sleeping modes turn out to be excited \textit{simultaneously}. Such a remarkable behaviour we have previously revealed for the stability properties of the FPU-$\alpha$ chain (see~\cite{FPU1}).

\section{Analytical method for stability analysis of nonlinear normal modes in the thermodynamic limit}

For studying stability of NNMs in the thermodynamic limit ($N\rightarrow\infty$), we use a method which is similar, in some sense, to that developed in~\cite{LeoLeo2007}. However, our method is more general and, as a consequence, it can be applied to all nonlinear normal modes in the FPU-$\beta$ chain for the case $N\rightarrow\infty$. Below, this method is described in detail.

The standard linear stability analysis of a given NNM leads us to Eq.~(\ref{eqch81}) representing $N\times N$ system of differential equations with periodic coefficients. Obviously, such straightforward way proves to be especially complicated when $N\rightarrow\infty$.

On the other hand, we can apply the general group-theoretical method~\cite{Zhukov} for splitting the system~(\ref{eqch81}) into some independent subsystems $L_j$ of small dimensionalities $n_j$ due to the symmetry properties of the considered NNM. In~\cite{FPU2}, it has been shown that for all NNMs in the FPU-$\beta$ chain these dimensionalities ($n_j$) do not exceed 3. This fact is extremely useful for stability studying and we have already used it in~\cite{FPU2} for numerical construction of the stability diagrams for NNMs in the FPU-$\beta$ model with $\beta>0$.

At this point it is appropriate to dwell on the physical cause of the stability loss of nonlinear normal modes. This loss of stability can be treated in terms of ``parametric interactions'' between a given excited NNM and other modes with zero amplitudes (``sleeping'' modes) (see~\cite{PhysD98}).

In the simplest case, such interaction can be described by the Mathieu equation. Indeed, studying stability of the $\pi$-mode in the FPU-$\beta$ chain, we deal with splitting of the $N$-dimensional variational system into individual \textit{scalar} equations
\begin{equation}\label{dob181}
\ddot\nu_j+4\left[1+\frac{12\beta}{N}\nu^2(t)\right]\nu_j\sin^2k=0,
\end{equation}
where $\nu_j=\nu_j(t)$ is a sleeping mode, while the solution $\nu(t)$ of the Duffing equation~(\ref{eqch72}) describes the time-evolution of the $\pi$-mode.

In some approximation, we can replace the exact function $\nu(t)$ by its first Fourier harmonic. After substitution of this approximate function into Eq.~(\ref{dob181}), the latter can be transformed to the standard form of the Mathieu equation, which possesses an infinite set of stable and unstable regions. Depending on the value of the considered NNM amplitude [the function $\nu(t)$] one can get into a region of unstable motion and then infinitesimal solution of the Mathieu equation begins exponentially increase in amplitude over time evolution. This means that the original NNM loses its stability and transforms into a two-dimensional bush of vibrational modes.

The above phenomenon represents a parametric resonance: the parameter in square brackets in Eq.~(\ref{dob181}) changes periodically that leads, under certain conditions, to excitation of the sleeping degree of freedom described by the function $\nu_j(t)$.

Let us note that different sleeping modes $\nu_j(t)$ turn out to be unequal in relation to excitation by the given NNM $\nu(t)$. Indeed, in Fig.~\ref{figStabDiag}a it can be seen that $\nu_j(t)$ with wavenumbers $k$ closer to $k_{\text{res}}=\pi/2$ prove to be excited earlier, i.e.\ for smaller amplitudes $A$ of the $\pi$-mode. The values $k_{\text{res}}$ for different NNMs in the limit $N\to\nobreak\infty$ can be revealed in Fig.~\ref{figStabDiag}: they correspond to the points that contact with the $k$-axis. We need to study vicinities of the points ($k=k_{\text{res}}$, $A=0$) for obtaining scaling of the stability thresholds in the thermodynamic limit ($N\to\nobreak\infty$).

In general case, parametric interaction with the original (active) NNM leads to exciting not only one sleeping mode, but a certain set of such modes (for the FPU-$\beta$ this set can consist of 1, 2 or 3 sleeping modes). For example, the variational system for NNM B$[a^4,iu]$ can be decomposed into two-dimensional subsystems $L_j^{(2)}$ [see Eq.~(\ref{eqch82}) and the discussion below this equation]. Each of these subsystems contains two sleeping modes $\nu_j(t)$ and $\nu_{j'}(t)$, where $j'=\frac{N}{2}-j$ [in terms of wavenumber (\ref{eqch190}), these modes are denoted by $\nu_k(t)$ and $\nu_{\frac{\pi}{2}-k}(t)$]. This means that when NNM $\nu(t)$ loses its stability the both sleeping modes $\nu_k(t)$ and $\nu_{\frac{\pi}{2}-k}(t)$ are excited \textit{simultaneously}.

Similarly, the variational system for the NNMs B$[a^3,i]$ and B$[a^3,iu]$ can be decomposed into three-dimensional independent subsystems $L_j^{(3)}$. This means that these nonlinear normal modes lose  their stability simultaneously relative to the three sleeping modes with wavenumbers $k$, $k+\frac{2\pi}{3}$, $k+\frac{4\pi}{3}$.

Since there are no analytical results on parametric resonance in such cases, we have to use the general Floquet stability analysis to obtain thresholds of the stability loss of the corresponding NNMs.

In terms of the Floquet method, the stability loss of a given NNM can be interpreted in the following manner. When we vary the amplitude $A$ of NNM in the stability region, Floquet multipliers move on the unit circle. Some of them can leave this circle after colliding with each other, and this phenomenon means that the considered NNM loses stability. On the other hand, Floquet multipliers represent eigenvalues of the monodromy matrix $\hat X(\pi)$, and they can be found as the roots $\lambda_j$ of its \textit{characteristic polynomial}.

A bifurcation from stability to instability of the considered NNM takes place when some roots of this polynomial coincide with each other. This fact can be revealed by vanishing of the \textit{discriminant} $D$ of the monodromy matrix characteristic polynomial. Indeed, in general case,
\begin{equation}\label{dop182}
D=\prod_{i<j}(\lambda_i-\lambda_j)^2
\end{equation}
and coincidence of every pair of eigenvalues $(\lambda_i,\lambda_j)$ leads to the relation $D=0$.

Now let us consider some technical details of our method for studying stability of NNMs in the thermodynamic limit.

The explicit form of the independent subsystems $L_j$ for all FPU-$\beta$ NNMs can be found in~\cite{FPU2}. In the present paper, we write $L_j$ for each NNM in the following matrix-vector form:
\begin{equation}\label{eqch100}
\ddot{\vec\mu}+\left[a\hat\omega^2+b\nu^2(t)\hat M\right]\vec\mu=0.
\end{equation}
Here $\hat\omega^2$ and $\hat M$ are $n_j\times n_j$ constant matrices depending on the harmonic frequencies of the variables coupled by Eq.~(\ref{eqch100}).

In Table~\ref{table20}, we present the matrices $\hat\omega^2$ and $\hat M$ for each of six NNMs permissible in the FPU-$\beta$ chain in terms of the frequencies $\tilde\omega_1=\sin\left(\frac{k}{2}\right)$, $\tilde\omega_2=\sin\left(\frac{k}{2}+\frac{\pi}{3}\right)$, $\tilde\omega_3=\sin\left(\frac{k}{2}+\frac{2\pi}{3}\right)$, $\tilde\omega_4=\sin\left(\frac{k}{2}+\frac{\pi}{2}\right)$.

\begin{table*}
\centering
\caption{Input data for stability analysis of nonlinear normal modes in the FPU-$\beta$ chain}\label{table20}
\begin{tabular}{|l|c|c|c|}
  \hline
  NNM          & Energy vs. NNM's amplitude             & $\hat\omega$                & $\hat M$ \\
  \hline
  B$[a^2,i]$   & $2A^2+\frac{4\beta}{N}A^4$             & $\tilde\omega_1$                    & $3\,\tilde\omega_1^2$  \\
  B$[a^3,i]$   & $\frac{3}{2}A^2+\frac{27\beta}{8N}A^4$ & $\frac{2}{\sqrt{3}}\left(\begin{array}{l}
                                                                                   \tilde\omega_1\\
                                                                                   \quad\tilde\omega_2\\
                                                                                   \quad\quad\tilde\omega_3\\
                                                                                   \end{array}\right)$ & $\frac{4}{3}\left(\begin{array}{ccc}
                                                                                                                2\tilde\omega_1^2&-\tilde\omega_1\tilde\omega_2&\tilde\omega_1\tilde\omega_3\\
                                                                                                                -\tilde\omega_1\tilde\omega_2&2\tilde\omega_2^2&-\tilde\omega_2\tilde\omega_3\\
                                                                                                                \tilde\omega_1\tilde\omega_3&-\tilde\omega_2\tilde\omega_3&2\tilde\omega_3^2\\
                                                                                                                \end{array}\right)$ \\
  B$[a^3,iu]$  & $\frac{3}{2}A^2+\frac{27\beta}{8N}A^4$ & $\frac{2}{\sqrt{3}}\left(\begin{array}{l}
                                                                                   \tilde\omega_1\\
                                                                                   \quad\tilde\omega_2\\
                                                                                   \quad\quad\tilde\omega_3\\
                                                                                   \end{array}\right)$ & $\frac{4}{3}\left(\begin{array}{ccc}
                                                                                                                2\tilde\omega_1^2&\tilde\omega_1\tilde\omega_2&-\tilde\omega_1\tilde\omega_3\\
                                                                                                                \tilde\omega_1\tilde\omega_2&2\tilde\omega_2^2&\tilde\omega_2\tilde\omega_3\\
                                                                                                                -\tilde\omega_1\tilde\omega_3&\tilde\omega_2\tilde\omega_3&2\tilde\omega_3\\
                                                                                                                \end{array}\right)$ \\
  B$[a^4,ai]$  & $A^2+\frac{\beta}{N}A^4$               & $\sqrt{2}\tilde\omega_1$            & $6\,\tilde\omega_1^2$  \\
  B$[a^4,iu]$  & $A^2+\frac{2\beta}{N}A^4$              & $\sqrt{2}\left(\begin{array}{cc}
                                                                         \tilde\omega_1&\\
                                                                         &\tilde\omega_4\\
                                                                         \end{array}\right)$ & $3\left(\begin{array}{cc}
                                                                                                       \tilde\omega_1^2&\tilde\omega_1\tilde\omega_4\\
                                                                                                       \tilde\omega_1\tilde\omega_4&\tilde\omega_4^2\\
                                                                                                       \end{array}\right)$ \\
  B$[a^3u,ai]$ & $\frac{1}{2}A^2+\frac{3\beta}{8N}A^4$  & $2\left(\begin{array}{l}
                                                                                   \tilde\omega_1\\
                                                                                   \quad\tilde\omega_2\\
                                                                                   \quad\quad\tilde\omega_3\\
                                                                  \end{array}\right)$ & $4\left(\begin{array}{ccc}
                                                                                               2\tilde\omega_1^2&-\tilde\omega_1\tilde\omega_2&\tilde\omega_1\tilde\omega_3\\
                                                                                               -\tilde\omega_1\tilde\omega_2&2\tilde\omega_2^2&-\tilde\omega_2\tilde\omega_3\\
                                                                                               \tilde\omega_1\tilde\omega_3&-\tilde\omega_2\tilde\omega_3&2\tilde\omega_3^2\\
                                                                                               \end{array}\right)$ \\
  \hline
\end{tabular}
\end{table*}

The time-depending function $\nu(t)$ entering Eq.~(\ref{eqch100}) is a periodic solution to the governing Eq.~(\ref{eqch38}). For every NNM in the FPU-$\beta$ chain, this governing equation represents Duffing equation
\begin{equation}\label{eqch39}
\ddot\nu+a\nu+b\nu^3=0,
\end{equation}
where $a$ is the squared frequency of the harmonic approximation, while $b={\frac{\beta}{N}\cdot\gamma}$ is a nonlinearity coefficient.

Eq.~(\ref{eqch39}) is called the \textit{hard} (\textit{soft}) Duffing equation if $b>0$ ($b<0$). For initial conditions $\nu(0)=A$, $\dot\nu(0)=0$, the solution of the hard Duffing equation can be written in the form
\begin{equation}\label{eqch40a}
\nu(t)=A\cn(\Omega t,\kappa^2).
\end{equation}
Here
\begin{equation}\label{eqch41a}
\Omega^2=a/(1-2\kappa^2),
\end{equation}
while modulus $\kappa$ of the Jacobi elliptic cosine is determined by the relation
\begin{equation}\label{eqch42}
2\kappa^2=bA^2/(a+bA^2).
\end{equation}
The solution~(\ref{eqch40a}) represents periodic function with the period
\begin{equation}\label{eqch43}
T=4K(\kappa)/\Omega,
\end{equation}
where $K(\kappa)$ is the complete elliptic integral of the first kind.

For the same initial conditions, the solution of the soft Duffing equation ($b<0$) can be written in the form
\begin{equation}\label{eqch50}
\nu(t)=A\sn(\Omega t+K(\kappa),\kappa^2),
\end{equation}
with
\begin{equation}\label{eqch51}
\Omega^2=a/(1+\kappa^2),
\end{equation}
\begin{equation}\label{eqch52}
\kappa^2=|b|A^2/(2a-|b|A^2),
\end{equation}
\begin{equation}\label{eqch53}
T=4K(\kappa)/\Omega.
\end{equation}

It is convenient to introduce the time scaling
\begin{equation}\label{eqch101}
\tau=\frac{2\pi}{T}t=\frac{\pi\Omega}{2K(\kappa)}t
\end{equation}
which transforms Eq.~(\ref{eqch100}) into the equation with $\pi$-periodic coefficients. As a result of this scaling the form of Eq.~(\ref{eqch100}) does not change, but the constant $a$ and $b$, entering this equation, must be multiplied by $4K^2(\kappa)/(\pi^2\Omega^2)$. We do not change notations in Eq.~(\ref{eqch100}), however, we imply below that the above transformations are already fulfilled.

Our further stability analysis of NNMs reduces to investigating the stability of the zero solution of Eq.~(\ref{eqch100}). The analysis consists of the following steps:

\textbf{Step 1.} Simplification of Eq.~(\ref{eqch100}) in the thermodynamic limit ($N\rightarrow\infty$). In this case $b\sim 1/N$ and we can decompose the coefficients of Eq.~(\ref{eqch100}) into power series with respect to the small dimensionless parameter
\begin{equation}\label{eqch102}
\eps=\frac{bA^2}{a}.
\end{equation}

\textbf{Step 2.} Search for the general solution of the approximate equation that was obtained as a result of Step~1.

\textbf{Step 3.} Construction of the monodromy matrix with the aid of the above solution.

\textbf{Step 4.} Construction of the characteristic polynomial of the monodromy matrix.

\textbf{Step 5.} Analyzing discriminant of the characteristic polynomial in the limit $N\to\nobreak\infty$.

Let us consider these steps in turn.

\subsection*{Step 1}

The function $\nu(t)$ in the form~(\ref{eqch40a}) for the case $\beta>0$ and in the form~(\ref{eqch50}) for $\beta<0$ must be substituted into Eq.~(\ref{eqch100}) taking into account that modulus $\kappa$ goes to zero when $N\rightarrow\infty$. To simplify $\nu(t)$, we use the following formulas from the theory of elliptic functions~\cite{AbramowitzStegun}:
\begin{equation}\label{eqch120}
\cn(u,\kappa^2)=\frac{2\pi}{\kappa K(\kappa)}\sum_{n=1}^\infty\frac{q^{\frac{2n-1}{2}}}{1+q^{2n-1}}\cos\frac{(2n-1)\pi u}{2K(\kappa)},
\end{equation}
\begin{equation}\label{eqch121}
\sn(u,\kappa^2)=\frac{2\pi}{\kappa K(\kappa)}\sum_{n=1}^\infty\frac{q^{\frac{2n-1}{2}}}{1-q^{2n-1}}\sin\frac{(2n-1)\pi u}{2K(\kappa)},
\end{equation}
where
\begin{align}
K(\kappa)=&\frac{\pi}{2}\left\{1+\left(\frac{1}{2}\right)^2\kappa^2+\left(\frac{1\cdot 3}{2\cdot 4}\right)^2\kappa^4+\right.\nonumber\\
&\phantom{\frac{\pi}{2}}\quad\left.\ldots+\left[\frac{(2n-1)!!}{2^n n!}\right]^2\kappa^{2n}+\ldots\right\}\label{eqch122}
\end{align}
is the complete elliptic integral of the first kind, while
\[
q=q(\kappa)=\exp\left(-\pi\frac{K(\kappa')}{K(\kappa)}\right)
\]
($\kappa'=\sqrt{1-\kappa^2}$ is the complimentary modulus of the elliptic functions). Note that the modulus~$\kappa$ depends on $bA^2$ in a different manner for the cases $\beta>0$ and $\beta<0$ [see Eqs.~(\ref{eqch42}) and (\ref{eqch52}), respectively].

Now, we have to decompose the left-hand-side of Eq.~(\ref{eqch100}) into the power series with respect to the small parameter $\eps=bA^2/a$. This very cumbersome decomposition has been performed with the aid of the Maple{\tiny\textsuperscript{\texttrademark}} mathematical package. The corresponding result can be written as follows:
\begin{widetext}
\begin{align}
\frac{d^2}{d\tau^2}\vec\mu+&\left\{\hat\omega^2\quad+\quad\left[-\frac{3}{4}\hat\omega^2+\left(\frac{1}{2}+\frac{1}{2}\cos 2\tau\right)\hat M\right]\eps+\right.\nonumber\\
&\quad\left[\frac{75}{128}\hat\omega^2+\left(-\frac{13}{32}-\frac{3}{8}\cos 2\tau+\frac{1}{32}\cos 4\tau\right)\hat M\right]\eps^2+\label{eqmy40}\\
&\quad\left.\left[-\frac{243}{512}\hat\omega^2+\left(\frac{87}{256}+\frac{597}{2048}\cos 2\tau-\frac{3}{64}\cos 4\tau+\frac{3}{2048}\cos 6\tau\right)\hat M\right]\eps^3\right\}\vec\mu+\nonumber\\
& O\left(\eps^4\right)=0,\nonumber
\end{align}
\end{widetext}

\subsection*{Step 2}

Eq.~(\ref{eqmy40}) represents a system of differential equations with time-periodic coefficients and to construct the corresponding monodromy matrix we must obtain its solution for $t=\pi$. On the other hand, for small time intervals, the solution of Eq.~(\ref{eqmy40}) can be found by a simple perturbation theory. To that end we decompose $\vec\mu(t)$ into a formal series
\begin{equation}\label{eqch200}
\vec\mu(t)=\sum_{n=0}^\infty\eps^n\vec\mu_n(t),
\end{equation}
substitute it into Eq.~(\ref{eqmy40}) and equate to zero the terms with every fixed power of the small parameter~$\eps$. As a result, we get the following set of differential equations
\begin{widetext}
\begin{subequations}\label{eqmy41}
\begin{align}
\ddot{\vec\mu}_0+\hat\omega^2\vec\mu_0=&\,0,\label{eqmy41a}\\
\ddot{\vec\mu}_1+\hat\omega^2\vec\mu_1=&-\left[-\frac{3}{4}\hat\omega^2+\left(\frac{1}{2}+\frac{1}{2}\cos 2\tau\right)\hat M\right]\vec\mu_0,\label{eqmy41b}\\
\ddot{\vec\mu}_2+\hat\omega^2\vec\mu_2=&-\left[-\frac{3}{4}\hat\omega^2+\left(\frac{1}{2}+\frac{1}{2}\cos 2\tau\right)\hat M\right]\vec\mu_1\label{eqmy41c}\\
&-\left[\frac{75}{128}\hat\omega^2+\left(-\frac{13}{32}-\frac{3}{8}\cos 2\tau+\frac{1}{32}\cos 4\tau\right)\hat M\right]\vec\mu_0,\nonumber\\
\ldots\nonumber
\end{align}
\end{subequations}
\end{widetext}

Because of the diagonal form of the matrix $\hat\omega^2$ (see Table~\ref{table20}), these equations determine certain sets of harmonic oscillators with different time-periodic external forces. Each of these oscillators is described by equation
\[
\ddot x+\omega^2 x=f(\tau).
\]
The general solution to this equation, obtained by the method of variation of arbitrary constants, can be written in the form
\begin{equation}\label{eqch201}
x(\tau)=c_1\sin\omega\tau+c_2\cos\omega\tau+\frac{1}{\omega}\int\limits_0^{\tau}f(t)\sin[\omega(\tau-t)]dt,
\end{equation}
where $f(t)$, in our case, represents a \textit{sum} of time-periodic functions whose frequencies are \textit{incommensurable}. Indeed, for the most NNMs from Table~\ref{table20}, $\hat\omega^2$ are matrices with different diagonal elements and, therefore, the components of the vector $\vec\mu_0(\tau)$ from~(\ref{eqmy41a}) vibrate with different frequencies. Substituting $\vec\mu_0(\tau)$ into~(\ref{eqmy41b}) leads to mixing its time-depended components because of multiplying by the matrix $\hat M$, and such a mixing produces more and more complicated terms in r.h.s.\ of Eqs.~(\ref{eqmy41}) when we take into account higher orders in the decomposition~(\ref{eqmy40})

\subsection*{Step 3}

The usual way to study stability of a given periodic dynamical regime is the Floquet method. In this method, we linearize nonlinear equations of motion in the vicinity of the periodic solution and calculate the \textit{monodromy matrix} $\hat X(T)$ by integrating $2n$ times the linearized equations with time-periodic coefficients over one period $T$ using specific initial conditions [$n$ is the number of equations in~(\ref{eqmy40})]. These conditions are determined by the successive columns of $2n\times 2n$ identity matrix.

Solving Eqs.~(\ref{eqmy41}) in step-by-step manner, we can construct the approximate analytical solution to Eq.~(\ref{eqmy40}) up to a fixed order of the small parameter $\eps$. With the aid of this solution, we are able to obtain the monodromy matrix $\hat X(\pi)$ for Eq.~(\ref{eqmy40}), where $\pi$ is the period of its coefficients.

The stability of the considered periodic solution is determined by Floquet multipliers representing eigenvalues of the monodromy matrix. If all these multipliers lie on the unit circle, the solution is \textit{linear stable}. In other case, the solution \textit{linear unstable}.

\subsection*{Step 4}

We obtain eigenvalues of the monodromic matrix $\hat X(\pi)$ as the roots of its characteristic polynomial.

Let us remind that according to the Newton formulas, the coefficients of the characteristic polynomial
\[
\det(\hat A-\lambda\hat E)\equiv(-1)^N(\lambda^N+p_1\lambda^{N-1}+\cdots+p_{N-1}\lambda+p_N)
\]
of any $N\times N$ matrix $\hat A$ can be expressed via the sums
\begin{equation}\label{eqch500}
s_k=\sum_{i=1}^N\lambda_i^k,\quad k=1,\ldots,N
\end{equation}
with the aid of the recurrence relation
\[
p_k=-(s_k+p_1s_{k-1}+p_2s_{k-2}+\cdots+p_{k-1}s_1)/k.
\]

Thus, we have
\begin{align*}
p_1 =& -s_1,\\
p_2 =& -(s_2+p_1s_1)/2,\\
p_3 =& -(s_3+p_1s_2+p_2s_1),\\
	 & \cdots
\end{align*}

On the other hand, all sums $s_k$ in Eq.~(\ref{eqch500}) can be found directly by means of traces of the matrix $\hat A$:
\[
s_k=\Tr(\hat A^k),\quad k=1,\ldots,N.
\]

It is well known, that in the case of any Hamiltonian system with $n$ degree of freedom Floquet multipliers $\lambda_i$ form pairs $\lambda_i,\lambda_i^{-1}$ ($i=1,\ldots,N$) and, as a consequence, the characteristic polynomial $f(\lambda)$ of the monodromic matrix $\hat X(\pi)$ proves to be \textit{palindromic}:
\begin{equation}\label{eqchA1}
f(\lambda)=\lambda^{2n}-a_1\lambda^{2n-1}+a_2\lambda^{2n-2}-\ldots+a_2\lambda^2-a_1\lambda+1=0.
\end{equation}
with the following coefficients $a_i$:
\begin{align*}
a_1=&\Tr\hat X(\pi),\\
a_2=&\frac{1}{2}\left[\Tr^2\hat X(\pi)-\Tr\hat X^2(\pi)\right],\\
    &\cdots
\end{align*}

Now we have to obtain formulas for discriminants $D_1$, $D_2$ and $D_3$ for the corresponding palindromic characteristic polynomials of $n=1,2,3$ degrees via traces of monodromy matrices.

From the above Newton formulas applied to the polynomial~(\ref{eqchA1}), we obtain the following formulas for its coefficients:
\begin{align}
a_1=&\Tr\hat X(\pi),\nonumber\\
a_2=&\frac{1}{2}\left[\Tr^2\hat X(\pi)-\Tr\hat X^2(\pi)\right],\label{eqchA2}\\
a_3=&\frac{1}{6}\left[\Tr^3\hat X(\pi)-3\Tr\hat X(\pi)\Tr\hat X^2(\pi)+2\Tr\hat X^3(\pi)\right].\nonumber
\end{align}

On the other hand, one can express discriminant $D_n$ explicitly via these coefficients of the characteristic polynomial. With the aid of Maple{\tiny\textsuperscript{\texttrademark}}, we finally find
\begin{align}
D_1=&(a_1+2)(a_1-2),\label{eqch300}\\
D_2=&(a_2+2a_1+2)(a_2-2a_1+2)(8+a_1^2-4a_2)^2,\label{eqch301}\\
D_3=&(a_3+2a_2+2a_1+2)(a_3-2a_2+2a_1-2)\cdot\nonumber\\
	&(9a_1^2+54a_1a_3-27a_3^2-42a_1^2a_2+18a_1a_2a_3\label{eqch302}\\
	&\,-4(a_2-3)^3+8a_1^4+a_1^2a_2^2-4a_1^3a_3)^2\nonumber
\end{align}
where $a_1$, $a_2$, $a_3$ are given by Eqs.~(\ref{eqchA2}).

Now we present some illustrations of the above discussed technique.

\textit{Example 1.} Nonlinear normal mode B$[a^2,i]$ ($\pi$-mode): $|x,-x|$.

Firstly, let us discuss the case $\beta>0$. One can see that there exist modes, corresponding to the left and right sides of the black region in Fig.~\ref{figStabDiag}a, which are not excited by parametric interactions with the $\pi$-mode. This fact was revealed analytically in~\cite{SanduskyPage1994} with the aid of Rotating Wave Approximation (RWA). In~\cite{Dauxois1997}, in the framework of the same approximation the following relation between the amplitude threshold value $A_c$ and the wavenumber $k$ (i.e.\ the boundary curve of the black region in Fig.~\ref{figStabDiag}a) was obtained:
\[
A_c=\frac{|\sin\frac{k-\pi}{2}|}{\sqrt{9\cos^2\frac{k-\pi}{2}-3}},
\]

This analytical formula is in good agreement with the numerical results.

Let us now consider the stability threshold of the $\pi$-mode in the thermodynamic limit $N\to\nobreak\infty$ using the above discussed method. We have to consider the vicinity of the point $(\frac{\pi}{2},0)$ on the $(k-A)$ plane ($k_{\text{res}}=\pi/2$). The one-dimensional constant matrices of the decoupled variational system can be found in Table~\ref{table20}:
\[
\hat\omega_k^2=\sin^2\frac{k}{2},\quad \hat M_k=3\sin^2\frac{k}{2}.
\]

The monodromy matrix also depends on the wavenumber $k$ and we can decompose its trace $\Tr\hat X(\pi)$ into the Taylor series in two small parameters $\Delta k=k-k_{\text{res}}$ and $\eps=bA^2/a$ (in our case, $a=4$, $b=16\beta/N$ and, therefore, $\eps=4\beta A^2/N$). This decomposition read
\begin{align*}
\Tr\hat X(\pi)=&\left(-2+\frac{1}{64}\pi^2\Delta k^4-\frac{1}{1536}\pi^2\Delta k^6+\ldots\right)+\\
&\left(-\frac{3}{32}\pi^2\Delta k^2+\frac{7}{512}\pi^2\Delta k^4+\ldots\right)\eps+\\
&\frac{27}{256}\pi^2\Delta k^2\eps^2+\ldots
\end{align*}

In the considered case, $n=1$ and the corresponding discriminant is
\[
D_1=(a_1+2)(a_1-2),
\]
where $a_1=\Tr\hat X(\pi)$.

The condition $D_1=0$ leads to the equations
\begin{equation}\label{eqch261}
a_1+2=0\quad\text{or}\quad a_1-2=0.
\end{equation}

From the first of these equation, we find
\begin{align*}
a_1+2=&\left(\frac{1}{64}\pi^2\Delta k^4-\frac{1}{1536}\pi^2\Delta k^6\right)+\\
&\left(-\frac{3}{32}\pi^2\Delta k^2+\frac{7}{512}\pi^2\Delta k^4\right)\eps+\\
&\frac{27}{256}\pi^2\Delta k^2\eps^2=0,
\end{align*}
and then we obtain
\[
\eps=\frac{1}{6}\Delta k^2+\frac{7}{144}\Delta k^4+\text{O}(\Delta k^6).
\]

Substitution of $\eps=\frac{4\beta}{N}A_c^2$ and $\Delta k=\frac{2\pi}{N}$ leads us to the following result
\[
A_c=\frac{\pi}{\sqrt{6\beta N}}+\frac{7\pi^3}{12N^2\sqrt{6\beta N}}+\text{O}(N^{-9/2}).
\]

The corresponding energy per one particle is
\[
\frac{E_c}{N}=\frac{1}{N}\left(\frac{aA_c^2}{2}+\frac{bA_c^4}{4}\right)=\frac{\pi^2}{3\beta N^2}+\frac{11\pi^4}{36\beta N^4}+\text{O}(N^{-6}).
\]

The second equation~(\ref{eqch261}) leads to a contradiction with the condition of smallness of the parameter $\eps$ and, therefore, this equation doesn't produce instability of the $\pi$-mode in the limit $N\to\nobreak\infty$.

Note that the analytical dependence $E_c/N=\pi^2/(3\beta N)$ was revealed in~\cite{BermanKolovskij1984}, and later was recovered in~\cite{Flach1996,PoggiRuffo1997}.

In the case $\beta<0$, the stability properties of the $\pi$-mode are utterly different. Indeed, one can see in Fig.~\ref{figStabDiag}a (right column) that this mode turns out to be stable up to the finite value of its amplitude $A_c$. Using a numerical method, we found that in the thermodynamic limit $N\to\nobreak\infty$ $A_c=0.393$ ($E_c=0.213$).

\textit{Example 2.} Nonlinear normal mode B$[a^4,iu]$: $|x,x,-x,-x|$.

The variational system for this NNM is decoupled into $2\times 2$ subsystems with time-periodic coefficients. As a result, for studying the stability loss of the mode B$[a^4,iu]$ we have to vanish the discriminant $D_2$ from~(\ref{eqch301}).

The first factor of $D_2$ near the resonance wavenumber $k_{\text{res}}=\pi/2$ is equal to
\begin{equation}\label{eqch400}
a_2+2a_1+2=\frac{9}{128}\eps^2\pi^4\Delta k^2+\frac{1}{16}\pi^4\Delta k^4+\ldots
\end{equation}
Being positive, this factor cannot lead to the condition $D_2=0$.

The second factor of $D_2$ from Eq.~(\ref{eqch301}) reads
\begin{equation}\label{eqch401}
a_2-2a_1+2=16+\text{O}(\eps^2,\Delta k^2)
\end{equation}
also doesn't vanish in the thermodynamic limit $N\to\nobreak\infty$.

Only the last factor of the discriminant~$D_2$
\begin{equation}\label{eqch402}
\frac{9}{64}\eps^2\pi^4\Delta k^2-\frac{1}{16}\pi^4\Delta k^6+\ldots
\end{equation}
can lead to fulfilment of the condition $D_2=0$. This yields
\[
\eps=\pm\frac{2}{3}\Delta k^2,
\]
and, therefore,
\begin{equation}\label{eqch403}
A_c=\frac{\sqrt{2}\pi}{\sqrt{3|\beta|N}},\quad\quad E_c=\frac{2\pi^2}{3|\beta|N^2}.
\end{equation}

Note that this NNM has been investigated quite enough in~\cite{LeoLeo2007}. Above, we simply reproduced the main result of this paper by our method.

\textit{Example 3.} Nonlinear normal modes B$[a^3,i]$: $|x,0,-x|$ and B$[a^3,iu]$: $|x,-2x,x|$.

The variational systems for these NNMs can be decoupled into $3\times 3$ independent subsystems whose matrices are presented in Table~\ref{table20}.

Now one has to vanish the discriminant $D_3$ from Eq.~(\ref{eqch302}). With the aid of our method, we get the following results.

For both modes B$[a^3,i]$ and B$[a^3,iu]$, we have obtained identical scaling for the case $N\to\nobreak\infty$:
\begin{equation}\label{eqch420}
A_c=\frac{2\pi}{3\sqrt{\beta N}},\quad\quad E_c=\frac{2\pi^2}{3\beta N^2}.
\end{equation}

Some numerical results on stability of the NNM B$[a^3,i]$ have been found in~\cite{Bountis2006}, but we don't know any results on stability properties of the NNM B$[a^3,iu]$.

In conclusion, let us consider another scenario of the stability loss of NNMs. Indeed, up to this point, we have discussed only the loss of stability associated with parametric interactions of a given NNM with other (linear) normal modes of the FPU-$\beta$ chain.

Some NNMs in the FPU-$\beta$ chain, when $\eps=bA^2/a\to\nobreak 0$, transform not into \textit{one} linear normal mode (LNM), but into a certain \textit{superposition} of such modes. For example, NNM B$[a^4,ai]$
\begin{equation}\label{eqch700}
\vec X(t)=\nu(t)\cdot\frac{2}{\sqrt{2N}}\{0,1,0,-1\,|\,0,1,0,-1\,|\ldots\}
\end{equation}
transforms, in the case $\eps\to\nobreak 0$, into the linear combination
\begin{equation}\label{eqch701}
\xi_1\vec\Psi_{N/4}(t)+\xi_2\vec\Psi_{3N/4}(t),\quad\xi_1=-\xi_2=\frac{1}{\sqrt{2}}
\end{equation}
of two linear normal modes
\begin{equation}\label{eqch702}
\begin{array}{rcl}
\vec\Psi_{N/4}(t)  &=& \vec c_{N/4}\cdot\cos(\omega t),\\
\vec\Psi_{3N/4}(t) &=& \vec c_{3N/4}\cdot\cos(\omega t),
\end{array}
\end{equation}
where
\begin{equation}\label{eqch703}
\begin{array}{rcl}
\vec c_{N/4}  &=& \frac{1}{\sqrt{N}}\{1,-1,-1,1\,|\,1,-1,-1,1\,|\ldots\},\\
\vec c_{3N/4} &=& \frac{1}{\sqrt{N}}\{-1,-1,1,1\,|\,-1,-1,1,1\,|\ldots\},\\
\omega^2      &=& 2.
\end{array}
\end{equation}

One can also say that NNM B$[a^4,ai]$ is the result of the continuation of the superposition~(\ref{eqch701}) with respect to the nonlinearity parameter~$\eps$ of the considered FPU-$\beta$ chain. We have to emphasize that the continuation of an \textit{arbitrary} linear combination of two above discussed LNMs doesn't represent an exact solution to nonlinear dynamical equations of the FPU-$\beta$ chain, while the superposition with $\xi_1=-\xi_2=\frac{1}{\sqrt{2}}$ produces an exact solution.

Then, the following question arises: If we will increase the parameter $\eps$ from zero, is it possible that the discussed NNM~(\ref{eqch700}) loses stability because of transformation into a \textit{two-dimensional bush}
\begin{equation}\label{eqch710}
\nu(t)\vec c_{N/4}+\mu(t)\vec c_{3N/4}
\end{equation}
with two \textit{different} functions of time $\nu(t)$, $\mu(t)$? In general, such a bush describes not periodic, but a \textit{quasiperiodic} dynamical regime in the FPU-$\beta$ chain. In contrast to the previous case, this stability loss scenario doesn't imply any extension of the vibrational modes set, but it means breaking the correlation~(\ref{eqch701}) between two LNMs, $\nu_{N/4}(t)\vec\Psi_{N/4}$ and $\nu_{3N/4}(t)\vec\Psi_{3N/4}$. We don't discuss this scenario of the stability loss in the present paper. However, our analysis allows us to assert that such scenario is inessential for all six symmetry-determined NNMs in the FPU-$\beta$ chain.

\section{Results and discussion}

With the aid of the above discussed method, we investigated scaling of stability thresholds in the thermodynamic limit $N\to\nobreak\infty$ for all possible in the FPU-$\beta$ chain nonlinear normal modes for both cases $\beta>0$ and $\beta<0$. The summary results are presented in Tables~\ref{table3},~\ref{table4}. Let us comment on these results.

\begin{table}
\centering
\caption{Asymptotic behavior of the critical parameters of NNMs in the thermodynamic limit for the FPU-$\beta$ chain with $\beta>0$.}\label{table3}
\[
\begin{array}{|l|c|c|c|}
\hline
\text{NNM}     & A_c                                        & E_c/N                     & \text{Refs.}\\
\hline
B[a^2,i]       & \frac{\pi}{\sqrt{6\beta N}}                & \frac{\pi^2}{3\beta N^2}  &\text{\cite{Budinsky1983,BermanKolovskij1984,SanduskyPage1994,Flach1996,Dauxois1997,PoggiRuffo1997,Bountis2006a,Yoshimura2004,Dauxois2005}}\\
B[a^3,i]       & \frac{2\pi}{3\sqrt{\beta N}}               & \frac{2\pi^2}{3\beta N^2} &\text{\cite{Bountis2006}}\\
B[a^4,ai]      & \sqrt{\frac{2\pi}{3\beta}}                 & \frac{2\pi}{3\beta N}     &\text{\cite{Bountis2006,Bountis2006a}}\\
B[a^3,iu]      & \frac{2\pi}{3\sqrt{\beta N}}               & \frac{2\pi^2}{3\beta N^2} &\\
B[a^4,iu]      & \frac{\sqrt{2}\pi}{\sqrt{3\beta N}}        & \frac{2\pi^2}{3\beta N^2} &\text{\cite{LeoLeo2007}}\\
B[a^6,ai,a^3u] & \frac{2\pi}{\sqrt{3\beta N}}               & \frac{2\pi^2}{3\beta N^2} &\\
\hline
\end{array}
\]
\end{table}

\begin{table}
\centering
\caption{Asymptotic behavior of the critical parameters of NNMs in the thermodynamic limit for the FPU-$\beta$ chain with $\beta<0$.}\label{table4}
\[
\begin{array}{|l|c|c|c|}
\hline
\text{NNM}     & A_c                                        & E_c/N                     & \text{Refs.}\\
\hline
B[a^4,ai]      & \sqrt{\frac{2\pi}{3|\beta|}}               & \frac{2\pi}{3|\beta|N}    &\\
B[a^4,iu]      & \frac{\sqrt{2}\pi}{\sqrt{3|\beta|N}}       & \frac{2\pi^2}{3|\beta|N^2}&\text{\cite{LeoLeo2007}}\\
B[a^6,ai,a^3u] & 0                                          & 0                         &\\
\hline
\end{array}
\]
\end{table}

In the first column of Tables~\ref{table3},~\ref{table4}, the symbols of NNMs are given. In two next columns, we present the scaling laws for $N\to\nobreak\infty$ of stability thresholds for the amplitude of each NNM, $A_c(N)$, and for its specific energy, $\eps_c(N)$ (the energy per one particle of the chain). In the last column of the above tables, we give references to the papers in which stability thresholds of the corresponding NNMs are also discussed.

\subsection{The case $\beta>0$}

Firstly, let us pay attention to the following interesting fact: four of six NNMs in the FPU-$\beta$ chain, namely,\footnote{Note that time-dependence of these modes, i.e.\ the function $\nu(t)$, obeys different Duffing equations [see Eqs.~(\ref{eqch38})-(\ref{eqch_65})].}
\begin{equation}\label{eqch499}
\begin{array}{lll}
\text{B}[a^3,i]   &=& \nu(t)\cdot\frac{3}{\sqrt{6N}}\{1,0,-1\,|\,1,0,-1\,|\,\ldots\},\\
\text{B}[a^3,iu]  &=& \nu(t)\cdot\frac{1}{\sqrt{2N}}\{1,-2,1\,|\,1,-2,1\,|\,\ldots\},\\
\text{B}[a^4,iu]  &=& \nu(t)\cdot\frac{1}{\sqrt{N}}\{1,-1,-1,1\,|\,1,-1,-1,1\,|\,\ldots\},\\
\text{B}[a^3u,ai] &=& \nu(t)\cdot\frac{3}{\sqrt{6N}}\{{\scriptstyle 0,1,1,0,-1,-1\,|\,0,1,1,0,-1,-1\,|\,\ldots}\},\\
\end{array}
\end{equation}
possess \textit{identical} scaling of the stability threshold in the limit $N\to\nobreak\infty$:
\begin{equation}\label{eqch_500}
\eps_c(N)=\frac{2\pi^2}{3\beta N^2}.
\end{equation}

The scaling of $\eps_c(N)$ for the $\pi$-mode
\begin{equation}
\text{B}[a^2,i] = \nu(t)\cdot\frac{1}{\sqrt{N}}\{1,-1\,|\,1,-1\,|\,\ldots\}
\end{equation}
is \textit{exactly} twice less than that determined by Eq.~(\ref{eqch_500}).

Only for the mode
\begin{equation}
\text{B}[a^4,ai] = \nu(t)\cdot\frac{2}{\sqrt{2N}}\{0,1,0,-1\,|\,0,1,0,-1\,|\,\ldots\}\\
\end{equation}
the scaling law of the stability threshold turns out to be cardinally different:
\begin{equation}\label{eqch502}
\eps_c(N)=\frac{2\pi}{3\beta N}.
\end{equation}

A qualitative difference in scaling between this mode and all other NNMs for the FPU-$\beta$ chain with $\beta>0$ can be seen in the left column of Fig.~\ref{figStabDiag}.

\subsection{The case $\beta<0$}

The stability properties of the same NNMs in the FPU-$\beta$ chain with $\beta<0$ prove to be essentially different, as one can see in the right column of Fig.~\ref{figStabDiag}.

Firstly, for three NNMs, namely, B$[a^2,i]$, B$[a^3,i]$ and B$[a^3,iu]$ the stability thresholds $\eps_c(N)$ don't tend to zero when $N\to\nobreak\infty$. These modes lose their stability for a certain \textit{finite} value $A_c$ of the NNM's amplitude. For the above listed nonlinear normal modes, these values are equal respectively to $0.213$, $0.092$, and $0.138$.

The fundamental difference between scaling of $\eps_c(N)$ for the modes B$[a^4,ai]$ and B$[a^4,iu]$ takes place in the case $\beta<0$, as well as for the above discussed case $\beta>0$: $\eps_c(N)\sim 1/N$ for B$[a^4,ai]$ and $\eps_c\sim 1/N^2$ for B$[a^4,iu]$ (see Table~\ref{table4}).

Studying of the stability threshold for the NNM B$[a^3u,ai]$ proves to be more difficult. In this case, the loss of stability is determined by the second scenario discussed at the beginning of the present section. Normalizing the variational equations, described dynamics of the vibrational modes with basis vectors $\vec\Psi_{N/2}$, $\frac{1}{\sqrt{2}}(\vec\Psi_{N/6}\pm\vec\Psi_{5N/6})$, simultaneously with the Duffing equation, corresponding to NNM B$[a^3u,ai]$, we infer that for $\beta<0$ there exists an exponential \textit{detuning} between the above modes for an arbitrary small amplitude of the investigated NNM. This means that $\eps_c(N)$ turns out to be equal to zero as indicated in Table~\ref{table4}.

\section{Summary}

In the present paper, a certain asymptotic technique for studying the stability loss of nonlinear normal modes in the FPU-$\beta$ chain in the thermodynamic limit $N\to\nobreak\infty$ is developed. Using this technique we were able to obtain the scaling laws of the stability threshold $\eps_c(N)$ for all six symmetry-determined NNMs that are possible in the FPU-$\beta$ chain for both cases $\beta>0$ and $\beta<0$.

The general method~\cite{Zhukov} for splitting the variational system for a given dynamical regime in a physical system with discrete symmetry into independent subsystems of small dimensions was applied for investigation of stability of NNMs in the FPU-$\beta$ chain. The above dimensions for the considered case turn out to be equal to 1, 2 and 3. This splitting allows us to construct numerically the stability diagrams (Fig.~\ref{figStabDiag}) that can help to reveal many interesting properties of nonlinear normal modes, such as qualitative behaviour of stability thresholds $\eps_c(N)$ in the thermodynamic limit $N\to\nobreak\infty$, the existence of the ``second stability threshold'' for some NNMs, existence of finite limits of $\eps_c(N)$ for certain modes, etc.

\end{document}